\documentclass[review]{elsarticle}

\journal{arXiv.org}

\usepackage{hyperref}
\usepackage[table,x11names]{xcolor}
\usepackage{threeparttablex} 
\usepackage{booktabs}        
\usepackage{longtable}
\usepackage{threeparttable}  
\usepackage{booktabs}
\usepackage{pdflscape}
\usepackage[nointegrals]{wasysym}
\usepackage{bm}
\usepackage{textcomp}
\usepackage{color,soul}
\usepackage{multirow}
\usepackage{hhline}
\usepackage[a4paper,
 total={170mm,257mm},
 left=20mm,
 top=20mm, showframe=false]{geometry}

\usepackage[]{graphicx}
\DeclareGraphicsExtensions{.pdf}

\newcommand{\R}[1]{\textcolor{black}{#1}}
\newcommand{\C}{\CIRCLE}
\newcommand{\CC}{\C\CIRCLE}
\newcommand{\CCC}{\CC\CIRCLE}
\newcommand{\CCCC}{\CCC\CIRCLE}
\newcommand{\CCCCC}{\CCCC\CIRCLE}

\newcommand{\ca}{\Circle}
\newcommand{\cc}{\ca\Circle}
\newcommand{\ccc}{\cc\Circle}
\newcommand{\cccc}{\ccc\Circle}
\newcommand{\ccccc}{\cccc\Circle}

\newcommand{\cl}{\LEFTcircle}
\newcommand{\cll}{\cl\LEFTcircle}
\newcommand{\clll}{\cll\LEFTcircle}
\newcommand{\cllll}{\clll\LEFTcircle}
\newcommand{\clllll}{\cllll\LEFTcircle}
\newcommand{\hcl}{\LEFTCIRCLE}

\definecolor{mygray}{gray}{0.9}

\begin{document}
\begin{frontmatter}

\title{A Systematic Literature Review of Automated Techniques for Functional GUI Testing of Mobile Applications}

\author[1]{Yauhen Leanidavich Arnatovich\corref{corauthor}}
\cortext[corauthor]{Corresponding author}
\ead{yauhen001@e.ntu.edu.sg}
\author[1]{Lipo Wang}

\address[1]{School of Electrical and Electronic Engineering, Nanyang Technological University, Singapore}

\begin{abstract}
\paragraph{\textbf{Context}} Multiple automated techniques have been proposed and developed for mobile application GUI testing aiming to improve \textit{effectiveness}, \textit{efficiency}, and \textit{practicality}. The \textit{effectiveness}, \textit{efficiency}, and \textit{practicality} are 3 fundamental characteristics which testing techniques are built upon, and need to be continuously improved to deliver useful solutions for researchers and practitioners, and community as a whole.
\paragraph{\textbf{Objective}} In this systematic review, we attempt to provide a broad picture of existing mobile testing tools by collating and analysing their conceptual, and also performance characteristics including an estimation of \textit{effectiveness}, \textit{efficiency}, and \textit{practicality}.
\paragraph{\textbf{Method}} To achieve our objective, we specify 3 primary, and 14 secondary review questions, and conducted an analysis of 25 primary studies. We first individually analyse each primary study, and next analyse the primary studies as a whole. We developed a review protocol which defines all the details of our systematic review.
\paragraph{\textbf{Results}} From \textit{effectiveness}, we conclude that testing techniques which implement model-checking, symbolic execution, constraint solving, and search-based test generation approach tend to be more effective than those implementing random test generation. From \textit{efficiency}, we conclude that testing techniques which implement code search-based testing approaches tend to be more efficient than those implementing GUI model-based. From \textit{practicality}, we conclude that the more effective a testing technique is, the less efficient it will be. 
\paragraph{\textbf{Conclusion}} For \textit{effectiveness}, we observe that the existing automated testing techniques are not effective enough, and currently they achieve nearly half of the desired level of effectiveness. For \textit{efficiency}, we observe that current automated testing techniques are not efficient enough. In general, they provide medium-to-low efficiency requiring more than 30 minutes per application. For \textit{practicality}, we observe that only nearly half of the existing mobile testing tools could be used in practice, while the others are not practical due to their low effectiveness and efficiency. 
\end{abstract}

\end{frontmatter}


\section{Introduction}
A \textit{systematic literature review} (in short ``systematic review'') is useful means of collating and summarizing the results of all existing research works which are relevant to a particular research question, or topic, or phenomenon of interest \cite{fink2013conducting5,petticrew2008systematic,kitchenham2004procedures7,Kitchenham8}. Systematic literature review is a form of \textit{secondary} study which uses a well-defined methodology to identify, analyse and interpret the results of the related research works in an unbiased manner, which also should be repeatable, at least to a certain degree. The research works used in systematic reviews  are called \textit{primary} studies. In fact, the most reliable evidence of knowledge can be seen from the primary studies which are aggregated on a particular topic into one single place, i.e., systematic reviews. Therefore, systematic literature reviews are the recommended form of aggregation for empirical studies \cite{fink2013conducting5,petticrew2008systematic,khan2011systematic4}. 

The original purpose for conducting systematic literature reviews is to support evidence-based medicine. In fact, software engineering-related research has relatively little empirical results  compared to medical one. Also, software engineering research methods are not as rigorous as those used in the medical domain. As such, Kitchenham adapted the medical guidelines for systematic reviews to software engineering \cite{kitchenham2004procedures7}. Later on, the initial software engineering guidelines were updated including  insights from sociology research \cite{Kitchenham8}.

A systematic literature review is a form of data collating and analysis, which requires considerably more efforts than a conventional literature review. First, systematic reviews should define a review protocol which specifies the \textit{review questions} to be addressed, and \textit{research methods} which are to be used to perform the review.\footnote{defining review protocol is necessary to reduce the possibility of researcher bias.} After specifying the review questions, systematic reviews uses the review protocol to describe its review process which includes research methods. In the review protocol, systematic reviews should specify \textit{search strategy}, \textit{study selection} process using inclusion and exclusion criteria of primary studies, \textit{quality assessment} criteria, \textit{data collection} and \textit{analysis}, and \textit{dissemination strategy}. 

For \textit{search strategy}, systematic reviews should specify search terms and resources which will be searched for primary studies. Resources may include digital libraries, specific journals, conference proceedings, gray literature, the Internet, and others. For \textit{study selection}, systematic reviews should specify inclusion and exclusion criteria which are used to determine which studies are included in, or excluded from, a systematic review. For \textit{quality assessment}, systematic reviews should specify quality questions to assess the data quality of the primary studies. For \textit{data collection} and \textit{analysis}, systematic reviews should specify how the information required from each primary study will be obtained, and how the data will be presented to help answer review questions. For \textit{dissemination strategy}, systematic reviews should specify how the results will be circulated to potentially interested parties.

Conventional systematic literature reviews  aggregate results related to a specific review question, e.g., ``\textit{Is  testing technique \textit{A} more effective at faults detection than  \textit{B} ?}'' However, there are two other types of systematic studies which complement systematic literature reviews such as \textit{systematic mapping} and \textit{tertiary} studies. Systematic mapping studies have more broad review questions, e.g., ``\textit{What is the current status of the topic of interest \textit{X} ?}'' \cite{KITCHENHAM2010792}. A \textit{systematic mapping} study allows to classify the primary studies in a specific topic area at a high level of granularity. This helps to identify areas for more primary studies to be conducted. A \textit{systematic tertiary} study can be performed in a domain where a number of systematic reviews already exist. It is seen as a review of secondary studies related to the same research question, which is a systematic review of systematic reviews, in order to answer even wider review questions. 

There are 3 main stages in a systematic review: \textit{planning}, \textit{conducting}, and \textit{reporting}. During the \textit{planning} phase, the systematic review should identify the need for a review, specify review questions, and develop review protocol. During the \textit{conducting} phase, the systematic review should specify search strategy, selection criteria, quality assessment criteria, and data extraction and synthesis. During the \textit{reporting} phase, the systematic review should specify dissemination strategy, and format of the final report.

In this systematic review, we focus on automated functional GUI testing of mobile applications (simply ``apps''). We specify 3 review questions regarding \textit{effectiveness}, \textit{efficiency}, and \textit{practicality} of the testing techniques. To the best of our search, we are the first who specify such review questions, and also in the topic of interested related to an automated GUI testing in mobile. In fact, mobile research is relatively new  area, so there may not be many primary studies which can be aggregated by systematic reviews in unbiased manner. So, during our search for the related systematic studies, we found 1 systematic literature review \cite{slr1},\footnote{this systematic literature review also includes mapping study as a part.} 4 systematic mapping studies \cite{slr1,map1,map2,map3}, and 1 survey \cite{surv1}, all of  which are conducted for mobile software testing. 

In this systematic review, for primary studies search, we use 11  digital libraries,  8 academic search engines, 66 individual journals, and 34 conferences. We search for the primary studies which are published between 2010 and 2018 years, inclusively. Using our search strategy, we found 3,639 primary studies in total. After applying our selection strategy, we obtained 47 primary studies which are relevant to an automated testing in mobile. Next, using our quality assessment criteria, we selected 25 primary studies as a final set for analysis, and excluded 22 primary studies.

This systematic review has been prepared using the suggested guidelines for performing systematic literature reviews in software engineering \cite{slrguide}. Following the suggested structure and contents of the final reports for systematic reviews, we organized our systematic review as follows.
In Section~\ref{sec:protocol}, we describe our review protocol used to conduct this systematic review. 
In Section~\ref{sec:background}, we give a background of the topic of interest including summary of previous systematic studies, motivation of this systematic review, and specifying review questions. 
In Section~\ref{sec:reviewmethods}, we conduct our systematic review. We specify search strategy and data sources, study selection process using inclusion and exclusion criteria, study quality assessment, and data extraction and synthesis. 
In Section~\ref{sec:results}, we discuss the results of the primary studies, their benefits, adverse effects and gaps. We also discuss possible variations of the results with effect of their applications on larger scales. 
In Section~\ref{sec:threats}, we discuss the validity of the results considering bias in our systematic review. 
In Section~\ref{sec:conclusion}, we summarize the results, and give recommendations to the researchers for possible improvements of existing techniques, and the future research directions. Also, for practitioners, we highlight  practical implications based on the results of this systematic review. 

\section{Review protocol}
\label{sec:protocol}
Our review protocol has been developed following the guidelines for performing systematic literature reviews in software engineering \cite{slrguide}.

\paragraph{Background} We first give a background, Section~\ref{sec:background}, where we introduce common concepts of event-driven software, graphical user interface and importance of its testing. We also, provide a summary of related previous systematic reviews (Section~\ref{subsec:previousss}), explain motivation of this systematic review, and specify review questions to be answered by this review (Section~\ref{subsec:motiv}). 

\paragraph{Data sources} We identify data sources to be searched, Section~\ref{subsec:datasrc}, where we introduce 11 major digital libraries (Section~\ref{subsubsec:dl}), 8 major academic search engines (Section~\ref{subsubsec:academicse}), 66 individual journals (Section~\ref{subsubsec:jour}, Table~\ref{tbl:1}), and 34 conferences (Section~\ref{subsubsec:conf}, Table~\ref{tbl:2}).

\paragraph{Search strategy} We describe search strategy, Section~\ref{subsec:searchstr}, where we identify search string to be used to search for primary studies. We search all our data sources for primary studies which are published between 2010 and 2018 inclusively, and related to graphical user interface testing in mobile. In total, our search strategy has found 3,639 primary studies (Table~\ref{tbl:3}).

\paragraph{Study selection strategy} We describe study selection process, Section~\ref{subsec:studysel}, where we determine which studies are included in, and excluded from, our systematic review. For study selection, we assigned 3 authors. If there have been any disagreements, they  were resolved during our group meeting with supervisor.

To apply \textit{general criteria} for inclusion and exclusion of primary studies (Section~\ref{subsubsec:selectionstr}, Table~\ref{tbl:4}), we equally distributed the found number of primary studies among 3 authors. After this filtering step, in total, we obtained 47 primary studies. Next, we performed quality assessments of 47 selected primary studies. For each data quality question, we assign weighting coefficient from [0...1] with step 0.1 depending on to which extent this particular study answers the quality question; `0' indicates poor data quality, i.e., study does not answer this particular data quality question at all, while `1' indicates excellent data quality for this particular data quality question. 

To apply \textit{quality criteria} for inclusion and exclusion of primary studies (Section~\ref{subsubsec:qualityassmnt}, Table~\ref{tbl:5}), we assigned all 47 primary studies to all 3 authors. Using our quality questions, each author assigned his/her own weighting coefficients to each quality question for each primary study.  Next, the assigned values were averaged to obtain the final weighting coefficient for each quality question. By summing the obtained average values, we computed a total weight for each primary study. Upon discussion in our research group, for inclusion of a primary study, we set a minimum threshold for the total weight (quality index) of the primary study. If the total weight of primary study is 2.50 (i.e., 50\% of the possible maximum 5.00) or more, we include the study in this systematic review, otherwise we consider that the study is of poor quality, and thus it should be excluded from consideration.  After this filtering step, in total (Table~\ref{tbl:6}), we included 25 (Table~\ref{tbl:11}), and excluded 22 (Table~\ref{tbl:15}) primary studies.  

\paragraph{Data extraction strategy} We describe data extraction process, Section~\ref{subsec:dataextr}, where we define how the information required from each primary study will be obtained. We developed 3 data forms to collect and tabulate the data in a way helping us to answer secondary and primary review questions of this systematic review. The example data forms are shown in Table~\ref{tbl:8}, Table~\ref{tbl:9}, and Table~\ref{tbl:10}.  

For  data extraction, we assigned 2 authors. Each author extracted data independently from all 25 primary studies. Next, the filled data forms were collected from the authors, and the provided data was compared. If there have been any conceptional mismatches, they were resolved during our discussion with the other 2 authors of this systematic review.

\paragraph{Data synthesis} We describe the extracted data, Section~\ref{subsec:datasynth}, where we tabulate the data in a way helping to answer our review questions (Table~\ref{tbl:12}, Table~\ref{tbl:13}, and Table~\ref{tbl:14}). Undertaking a descriptive synthesis, we do not perform any meta-analysis of primary studies. We collate and summarize the extracted data using ``Line of argument synthesis'' approach \cite{noblit1988meta}, where we first individually analyse each primary study, and next analyse the primary studies as a whole.

\paragraph{Dissemination strategy} We provide the results, Section~\ref{sec:results}, where we discuss the findings of the primary studies. We report our results in 2 formats: (1) as a journal paper, and (2) as a chapter of a PhD thesis. 

\section{Background}
\label{sec:background}
A Graphical User Interface (GUI) is acknowledged as a crucial component of an event-driven software (e.g., mobile apps) \cite{5401169,chaudhary2016metrics,6571622}. In the event-driven software, e.g., real-world apps, the app GUI usually contain hundreds or even thousands of  elements \cite{mckain2008graphical}. According to \cite{637386,memon2002gui,Myers1995}, a large part of the app code is dedicated to the user interface so its testing is an essential part of the software development life cycle (SDLC),\footnote{software development life cycle is the process of dividing development work into several phases to improve design, product, and project management in a cycle.}  which may significantly improve the overall quality of software \cite{Harrold2000,mcconnell1996}. During testing phase, GUI can be tested by  executing each event individually and observing its behaviour \cite{908959}. However, it is not a trivial task since the behaviour of an event handler may depend on GUI internal state, the state of other entities (objects, event handlers)
and the external environment. Furthermore, the outcome of an event handler execution may vary depending on the concrete sequence of preceding events. As a result, each GUI event needs to be tested in a context of different states via generating and executing various sequences of GUI events \cite{acteve36,Yuan2007}. 

The impetus of GUI is to simplify a user interaction with the app. GUI takes user actions (e.g., touches, selections, typing etc.) as input, and changes the state of its GUI elements by translating the user actions into the platform-specific event handlers to execute corresponding app functionality. Providing such ``event-handler architecture'', event handlers may be created and maintained fairly independently so that complex software may be built using these loosely coupled pieces of code while offering many degrees of usage freedom via its GUI (e.g., users may choose to perform a particular task in different ways in terms of possible user actions, their number and execution order).

Modern mobile apps have a highly interactive nature and complex GUI structure. As such, automated GUI testing of mobile apps is a daunting task for the developers and testers. In fact, often the GUI testing is done manually where all possible combinations of the GUI elements for a given app screen are manually tested for functional correctness and aesthetic quality. The manual GUI testing is no doubt an effective approach, however it is inefficient, i.e., time-consuming, error-prone, and usually not complete, especially for a large software with complex GUIs. So, to facilitate manual testing,   various automated testing techniques have been introduced such as model-based testing \cite{utting,utting2010practical,sof207,broy2005model,1553582,Dalal1999}, concolic testing \cite{Sen2007,acteve19,paquesymbolic}, search-based testing \cite{baldoni2016survey,5954405}, evolutionary testing \cite{WEGENER2001841,evodroid37}, and combinatorial testing \cite{autodroidcomb}. Depending on which testing technique is implemented, the GUI testing can be performed via dynamic  or static  program analysis methods, or their combination.

\subsection{Previous systematic studies}
\label{subsec:previousss}
For the related systematic studies which are conducted for mobile software testing, we found 1 systematic literature review \cite{slr1}, 4 systematic mapping studies \cite{slr1,map1,map2,map3}, and 1 survey \cite{surv1}. We describe them in a sequential order starting from the systematic literature review, next mapping studies, and at last the survey.

\paragraph{Systematic mapping and literature review} This systematic study \cite{slr1} was conducted by 3 authors in 2015. The work is entitled ``\textit{Automated Testing of Mobile Applications: A Systematic Map and Review}'', and comprises 15 pages. Note that this systematic study combines mapping and literature review. The authors conducted the systematic study to identify and collate evidence about current state of research in an automated GUI testing of mobile apps. They identified and characterized automated testing approaches and techniques, and investigated major challenges via analysis and synthesis of the selected primary studies. 

To drive the systematic study, the authors specified 3 \textit{review questions}, and 3 \textit{mapping questions}. 
\begin{enumerate}
\item \textit{review}: What are the challenges of automated testing of mobile applications?
\item \textit{review}: What are the different approaches for automated testing of mobile applications? 
\item \textit{review}: What is the most used experimental method for evaluating automated testing of mobile applications? 
\item \textit{mapping}: Which are the main journals and conferences for automated testing of mobile applications?
\item \textit{mapping}: Which are the main authors for automated testing techniques research?
\item \textit{mapping}: How is the frequency of papers distributed according to their testing approach?
\end{enumerate}

In total, 83 primary studies were selected for analysis. The authors tabulated and synthesized the results in a way helping practitioners to provide recommendations about automated testing of mobile apps. The popularity of the main approaches was identified: model-based testing (30\%), capture/replay (15.5\%), model-learning testing (10\%), systematic testing (7.5\%), fuzz testing (7.5\%), random testing (5\%) and scripted based testing (2.5\%). The authors conclude that the number of the proposed approaches and techniques for automated testing of mobile apps has increased. They also highlight that in 40\% of the selected primary studies, the automated testing techniques use GUI-based models of the target apps. 

\paragraph{Systematic mapping study} This systematic mapping \cite{map3} was conducted by 3 authors in 2015--2016 years. The work is entitled ``\textit{A systematic mapping study of mobile application testing techniques}'', and comprises 23 pages. The authors conducted the systematic mapping to categorize and structure the research evidence which has been available so far in the area of mobile apps testing including their approaches, techniques and challenges they addressed.

To drive the systematic mapping study, the authors specified 1 primary, and 3 secondary \textit{research questions}:
\begin{enumerate}
\item What are the studies that empirically investigate mobile and smart phone application testing techniques and challenges?
\begin{enumerate}
\item What research approaches do these studies apply and what contribution facets do they provide? 
\item What kind of applications (industrial or simple) do these studies use in order to evaluate their solutions? 
\item Which journals and conferences included papers on mobile application testing? 
\end{enumerate}
\end{enumerate} 

In total, 79 primary studies were selected and classified. The authors identified several research gaps, and key testing issues which could be interesting to practitioners. They report that only  few studies do investigation on real-world mobile environments, and focus on eliciting testing requirements in the requirements engineering phase. The authors also highlight that there is no clear guidance for practitioners how to choose an automated testing tool, or  testing technique from variety of available ones. They authors suggest that there is a need for a clear road-map to guide the practitioners, and for researchers, there is need for more studies which address the issues of conformance to life cycle models, mobile services and mobile testing metrics. 

\paragraph{Systematic mapping study} This systematic mapping \cite{map2} was conducted by 3 authors in 2015. The work is entitled ``\textit{Mobile Application Verification: A Systematic Mapping Study}'', and comprises 17 pages. The authors conducted the systematic mapping focusing on software verification aspect of the mobile applications. They found definitive metrics and research evidence about mobile application testing, which could be helpful for researchers to identify possible gaps and new research directions.

To drive the systematic mapping study, the authors specified 5 \textit{research questions}:
\begin{enumerate}
\item What are the most frequently used test types for mobile applications? (Compatibility, Concurrent, Conformance, Performance, Security, Usability)?
\item Which research issues in mobile application testing are addressed and how many papers cover the different research issues? (Test Execution Automation, Test Case Generation, Test Environment Management, Testing on Cloud, Model Based Testing)? 
\item At what test level have researchers' studies most frequently? (Unit, Component, Integration, System, Acceptance)? 
\item What is the paper-publication frequency? 
\item Which journals include papers on mobile application testing? 
\end{enumerate}

In total, 123 primary studies were selected and classified. The authors summarized the studies which are conducted for mobile app testing, and performed a gap analysis to provide a map of state-of-the-art in automated testing of mobile apps. They conclude that mobile software testing research is open to new contributions. In particular, research on performance testing  may provide more further opportunities since there is a lack of studies in this area. The results also indicate immerging research needs in mobile app testing on the cloud to deal with test execution automation, or test environment management for system level functional testing. 

\paragraph{Systematic mapping study} This systematic mapping \cite{map1} was conducted by 2 authors in 2016. The work is entitled ``\textit{Quality Assurance of Mobile Applications: A Systematic Mapping Study}'', and comprises 13 pages. The authors conducted the systematic mapping to identify approaches which address the issue of quality assurance for mobile applications. They describe approaches based on a test level focus and quality, and addressed research challenges. 

To drive the systematic mapping study, the authors specified 7 \textit{research questions}:
\begin{enumerate}
\item Which testing types of quality assurance approaches exist?
\item Which testing levels are addressed? 
\item Which testing phases are addressed? 
\item Which qualities are addressed? 
\item Which kinds of automation are implemented? 
\item How are the approaches evaluated? 
\item Which challenges exist?
\end{enumerate}

In total, 230 primary studies were selected and classified. The authors found that mainly system testing is considered, while functional and non-functional properties are addressed during quality assurance with a slightly stronger focus on the former. They also highlight that automation of the testing process plays an important role for mobile-specific quality assurance, especially on GUI level, however, the maturity of such tools is low. For researchers, the results can help to identify further research directions, and motivate to  perform more accurate evaluation of the proposed approaches, especially for industrial cases. 

\paragraph{Survey} This survey \cite{surv1} was conducted by 3 authors in 2009--2011 years. The work is entitled ``\textit{Obstacles and opportunities in deploying model-based GUI testing of mobile software: a survey}'', and comprises 29 pages. The authors conducted this survey to identify possible obstacles and opportunities towards wider deployment of model-based testing approach in industry. The survey results indicate that even not being widely used in industry yet, model-based testing imposes a great interest among mobile software testing professionals. However, there is a need for further research to understand how to efficiently manage model construction during testing since larger models are often impractical. In addition, uniform metrics of tests effectiveness should be developed to enable comparison between different testing approaches. They also highlight that more research attention should be dedicated to developing testing techniques which can do a quick bug localization. 

\subsection{Motivation and review questions}
\label{subsec:motiv}
\paragraph{Motivation} Since beginning of mobile era, various techniques have been proposed and developed for mobile app GUI testing. Some of them are fully automated, while others still rely on the user inputs to a certain extent, e.g., semi-automated and manual. Nevertheless, many developed testing techniques have resulted in the testing tool (executable artefact).  In fact, all these tools aim to increase test coverage, optimize model construction, and eventually deliver a solution which can be used in practice. Increasing coverage is explained by a fact that the higher test coverage, the more app functionality is tested, thus enabling the testing tool to potentially discover more app bugs and failures. Also, the more optimal model can be constructed, the more efficient testing tool will be. This aspect is the same critical as the test coverage because modern apps usually have complex code and GUI structures innately yielding large or extremely large models. However, such large models unlikely to be fully (in systematic manner) explored resulting in lower test coverage, otherwise exploration time will grow exponentially and could even be infinite which is impractical. 

Test coverage is de facto useful means of showing effectiveness, while optimal model construction shows efficiency of the testing tools. In turn, practicality tightly depends on the effectiveness and efficiency, however it could also be decided up on individual estimation of effectiveness or efficiency. Effectiveness, efficiency, and practicality are 3 fundamental characteristics which testing tools are built upon, and aiming to continuously improve them in order to deliver useful solutions for researchers and practitioners, and community as a whole. To the best of our knowledge, there is no systematic review conducted in the field, which attempts to provide a broad picture of the existing mobile testing tools by collating and analysing their conceptual, and also performance characteristics including an estimation of \textit{effectiveness}, \textit{efficiency}, and \textit{practicality}. Therefore, in this systematic review, we decided to take this challenge, and give an attempt to evaluate various testing tools and their characteristics for automated functional GUI testing of mobile apps. 

\paragraph{Review questions}
Specifying the review questions (RQs) is the most important part of any systematic review as they drive the entire systematic review methodology. In fact, the critical issue in any systematic review is to ask the right question(s). As such, in this section, we identify primary and secondary RQs which are meaningful and important to researchers, and could also be valuable to  practitioners.

To drive our systematic review, we specify 3 primary, and \R{14} secondary RQs.
The primary RQs are as follows:
\begin{enumerate}
    \item[] RQ\#1 How effective the proposed GUI testing techniques are in mobile?
    \item[] RQ\#2 How efficient the proposed GUI testing techniques are in mobile?
    \item[] RQ\#3 How practical the proposed GUI testing techniques are in mobile?
\end{enumerate} 

To facilitate RQ\#1, we specify \R{6} secondary RQs:
\begin{enumerate}
    \item[] RQ\#1.1 What model paradigm is used? 
    \item[] RQ\#1.2 What test generation approach is used? 
    \item[] RQ\#1.3 What test generation criteria is used?
    \item[] RQ\#1.4 What textual input generation mechanism is used? 
    \item[] RQ\#1.5 What code coverage metric is used?
    \item[] RQ\#1.6 What code coverage results are achieved on average?
\end{enumerate}  

To facilitate RQ\#2, we specify \R{6} secondary RQs:
\begin{enumerate}
    \item[] RQ\#2.1 What testing approach is used? 
    \item[] RQ\#2.2 What testing technique is used? 
    \item[] RQ\#2.3 What search algorithm is used? 
    \item[] RQ\#2.4 What termination condition is used?
    \item[] RQ\#2.5 What app benchmark size is used?
    \item[] RQ\#2.6 What execution time is taken per app?
\end{enumerate}  

To facilitate RQ\#3, we specify \R{2} secondary RQs:
\begin{enumerate}
    \item[] RQ\#3.1 How effectiveness  impacts practicality? 
    \item[] RQ\#3.2 How efficiency impacts practicality? 
\end{enumerate}

\section{Review Methods}
\label{sec:reviewmethods}
In this section, we conduct a systematic literature review of the selected primary studies. We identify data sources and specify a search strategy, perform study selection and quality assessment, data extraction and synthesis in accordance with our developed review protocol (see Section~\ref{sec:protocol}). 

\subsection{Data sources}
\label{subsec:datasrc}
The aim of a systematic review is to find as many primary studies relating to the research question as possible using an unbiased search strategy. Therefore, the rigorous search process is a crucial and  identifying factor for the systematic reviews unlike traditional ones. The first step for searching  primary studies can be undertaken using digital libraries, however, in practice, it is not sufficient for a complete systematic review. As such, other relevant sources must also be searched, e.g., reference lists from relevant primary studies and review articles, research and industrial (company) journals, grey literature (i.e., technical reports, white papers, unpublished work, and work in progress), conference proceedings, and the Internet in overall. Also, using various sources for the primary studies search, helps to mitigate a problem in systematic reviews, which is know as \textit{publication bias} leading to systematic bias in systematic reviews. The \textit{publication bias} is the problem where positive results are more likely to be published rather than negative ones \R{\cite{slrguide}}. To address the issue with \textit{publication bias}, we perform an exhaustive search for primary studies. 

As suggested by Brereton \cite{BRERETON2007571}, there is no single source which can find all the primary studies. Thus, we identify multiple data sources including various digital libraries, academic search engines, individual journals, and conferences. Below, we introduce the selected relevant software engineering digital libraries, academic search engines, journals, and conferences. In this systematic review, the listed data sources are used for the search of primary studies. 

\subsubsection{Digital libraries}
\label{subsubsec:dl}
For digital libraries, we identify \R{11} major data sources which cover software engineering domain, and are relevant to software engineers. We define \textit{digital library} as a single source where  electronic articles can be searched, and downloaded as a full text. The selected digital libraries are listed below. 

\begin{enumerate}
\item Research at Google\footnote{research.google.com/pubs/papers.html}
\item IBM Research\footnote{domino.research.ibm.com/library/cyberdig.nsf}
\item IEEE Xplore\footnote{ieeexplore.ieee.org/Xplore/home.jsp}
\item ACM Digital Library (ACM DL)\footnote{dl.acm.org}
\item Wiley Online Library\footnote{onlinelibrary.wiley.com}
\item SpringerLink\footnote{link.springer.com}
\item ScienceDirect\footnote{www.sciencedirect.com}
\item JSTOR\footnote{www.jstor.org}
\item ResearchGate\footnote{www.researchgate.net}
\item arXiv\footnote{arxiv.org}
\item Academia.edu\footnote{www.academia.edu}
\end{enumerate}

\subsubsection{Academic search engines}
\label{subsubsec:academicse}
For the academic search engines (or simply ``search engines''), we identify \R{8} major search sources. We define \textit{search engine} as a single interface where electronic articles can only be  searched, while providing a link to an external source from where the found article (full text) can be downloaded. The selected search engines are listed below.

\begin{enumerate}
\item Google Scholar\footnote{scholar.google.com}
\item Microsoft Academic (MA)\footnote{academic.microsoft.com}
\item Ei Compendex\footnote{www.engineeringvillage.com}
\item Scopus\footnote{www.elsevier.com/scopus}
\item Web of Science\footnote{www.webofknowledge.com}
\item Inspec\footnote{theiet.org/inspec}
\item CiteSeerX\footnote{citeseer.ist.psu.edu}
\item dblp (Digital Bibliography \& Library Project)\footnote{dblp.org}
\end{enumerate}

Apart from the digital libraries, and search engines, in our systematic review, we also do a search for primary studies in the individual journals and conference proceedings, thus we justify our selection criteria as follows. We select all potential journals and conferences, aims and scope of which, are in software engineering domain including software quality, testing, validation, verification, and reliability.

\subsubsection{Journals}
\label{subsubsec:jour}
To select journals, we do a manual search in the database \textit{Master Journal List} from Clarivate Analytics,\footnote{mjl.clarivate.com} which includes all journal titles covered by Web of Science. In particular, we use specific search terms such as ``\textit{computer}'', ``\textit{information}'', ``\textit{technology}'', ``\textit{software}'', ``\textit{quality}'', ``\textit{testing}'', ``\textit{verification}'', ``\textit{validation}'', and ``\textit{reliability}'' to find journals which include these words in their titles. Note that Clarivate Analytics engine searches for exact string matches of the search terms in the journal titles, so the search terms should not be combined. After performing search using each of the individual search terms, we obtained \R{848} matching journals. Next, we manually verified relevance each of the journals by going through lists of their titles, and, if necessary, we also did a quick review of their aims and scope. If their titles are too generic, and aims and scope are not stated clearly, we searched articles in the journals with keywords which are specific to our domain. If the search produced results, we reviewed abstracts and conclusions of several found articles of the target journal to confirm its relevance, otherwise, we concluded that the journal was irrelevant. So, after removing irrelevant journals, and duplicates, we obtained \R{66} relevant journals. In Table~\ref{tbl:1}, we show the search results\footnote{searched in January, 2018} for the journals using the above-specified search criteria.

\begin{table}[!t]
\centering
\caption{\textbf{Search results for journals in Master Journal List from Clarivate Analytics.}
\label{tbl:1}}
\begin{tabular}{l|l|l|l} \hline
\textbf{No.} & \textbf{Search term in journals title} & \textbf{Total found journals} & \textbf{Total relevant journals} \\ \hline
1 & computer & 123 & 25 \\ \hline
2 & information & 212  & 29 \\ \hline
3 & technology & 411 & 7 \\ \hline
4 & software & 27  & 15 \\ \hline
5 & quality & 46  & 1 \\ \hline
6 & testing & 15  & 1 \\ \hline
7 & verification & 1 & 1 \\ \hline
8 & validation & 1  & 0 \\ \hline
9 & reliability & 12  & 1 \\  \hhline{=|=|=|=}
\multicolumn{2}{c|}{\textbf{Total journals}} & \textbf{848} & \textbf{80} \\ \hline
\multicolumn{2}{c|}{\textbf{Total journals excluding duplicates}} & \textbf{789} & \textbf{66} \\ \hline
\end{tabular}
\end{table}

\subsubsection{Conferences}
\label{subsubsec:conf}
To select conferences, we do a manual search \textit{CORE2018} database from CORE Conference Portal\footnote{portal.core.edu.au/conf-ranks} which provides an information about a collection of Computer Science conferences. In particular, we use specific search terms such as ``\textit{computer}'', ``\textit{information}'', ``\textit{technology}'', ``\textit{software}'', ``\textit{quality}'', ``\textit{test}'', ``\textit{verification}'', ``\textit{validation}'', and ``\textit{reliability}'' to find conferences which include these words in their titles. Note that CORE Conference Portal engine searches for exact or partial string matches of the search terms in the conference titles. Next, we select those which are recognized as flagship, excellent, or good software engineering conferences. We assume that research works which are published on such venues are likely with high quality of the conducted research and reported research results in comparison with other software engineering conferences. We identify flagship, excellent, and good software engineering conferences using the CORE Rankings Portal.\footnote{core.edu.au/conference-portal}
\begin{itemize}
\item A* -- a flagship conference, a leading venue in a discipline area.
\item A  -- an excellent conference, a highly respected venue in a discipline area.
\item B  -- a good conference, a well regarded venue in a discipline area.
\end{itemize}
After performing search using each of the individual search terms, and filtering the matching   conferences by their rank, we obtained \R{231} matching conferences with ranks A*, A, and B. Next, we manually verified relevance each of the conferences  by going through lists of their titles, and, if necessary, we also did review of their primary topics of interest which are listed on the home web-pages. If their titles are too generic, and topics do not indicate relevance to our systematic review, we searched articles in the conferences proceedings with keywords which are specific to our domain. If the search produced results, we reviewed abstracts and conclusions of several found articles from the latest proceedings of the target conference to confirm its relevance, otherwise, we concluded that the conference was irrelevant. So, after removing irrelevant conference, and duplicates, we obtained \R{34} relevant conferences. In Table~\ref{tbl:2}, we show the search results\footnote{searched in January, 2018} for the conferences using the above-specified search criteria.

\begin{table}[!t]
\centering
\caption{\textbf{Search results for A*, A, and B-rank conferences in CORE2018 database from CORE Conference Portal.}
\label{tbl:2}}
\begin{tabular}{l|l|l|l} \hline
\textbf{No.} & \textbf{Search term in conferences title} & \textbf{Total found conferences} & \textbf{Total relevant conferences} \\ \hline
1 & computer & 77 & 9 \\ \hline
2 & information & 59  & 5 \\ \hline
3 & technology & 26 & 4 \\ \hline
4 & software & 48  & 17 \\ \hline
5 & quality & 4  & 2 \\ \hline
6 & test &  8 & 3 \\ \hline
7 & verification & 5 & 4 \\ \hline
8 & validation & 1  & 1 \\ \hline
9 & reliability & 3  & 3 \\  \hhline{=|=|=|=}
\multicolumn{2}{c|}{\textbf{Total conferences}} & \textbf{231} & \textbf{48} \\ \hline
\multicolumn{2}{c|}{\textbf{Total conferences excluding duplicates}} & \textbf{203} & \textbf{34} \\ \hline
\end{tabular}
\end{table}

\subsection{Search strategy}
\label{subsec:searchstr}
The process of performing a systematic literature review must be transparent and replicable. Therefore, the search process must be documented in sufficient detail so that the readers will be able to assess its thoroughness \R{\cite{slrguide}}. We generate our search strategy which identifies existing systematic reviews, mapping, and potentially relevant primary studies. In particular, we prepared sophisticated search query using Boolean ANDs, ORs, and exact phrase matching, where search query words occur anywhere in the article. Also, for all the data sources, we search (if the option is available in the search source) for the articles which are dated between 2010 and 2018 inclusively. We justify low boundary (i.e., 2010) by the fact that the first releases of the touch-screen-based operating systems (e.g., Android and iOS) have become available to the wide public from the mid-late of 2010s. To construct our search query, we use field-specific, and closely related to the topic of interest keywords which directly target the research area covering in this systematic review. Therefore, we construct a search query which is neither too generic, nor too narrow. As such, we believe that our search strategy finds most of the potentially relevant primary studies, while filtering out irrelevant ones. 

Due to the relatively large number of the data sources used in this systematic review, we give an example how we constructed our search query using an instance of Google Scholar search engine. For the other data sources, we apply the same principals of the search strategy using ``Advanced search'' or ``Expert search'' option (if available in the search source) with only possible variations in the syntax, and/or search terms (it depends on the search source) of the search query, while preserving the same semantics of the query to ensure the same quality of the search. For example, in Google Scholar, we use ``Advanced search'' option where we construct our search query as follows \textbf{mobile graphical OR user OR interface OR box OR functional OR android OR ios OR execution OR systematic OR random OR symbolic OR concolic OR model OR online OR reliability OR verification "gui testing"}. Using this search query, the Google Scholar search engine finds articles (1) with \textit{all} of the words \textbf{mobile}, (2) with the \textit{exact phrase} \textbf{"gui testing"}, and (3) with \textit{at least one} of the words \textbf{graphical OR user OR interface OR box OR functional OR android OR ios OR execution OR systematic OR random OR symbolic OR concolic OR model OR online OR reliability OR verification}, which may occur anywhere in the articles which are dated between 2010 and 2018 inclusively.

Here we justify our choice of such search strategy. First, we require the word \textbf{mobile} to be in the article since focus of this systematic review is on mobile testing techniques. Second, we require the exact phrase \textbf{"gui testing"} to be in the article. We piloted our search strategy, and identified that this particular phase is very likely to be used by the authors if their articles are related to the testing of GUIs. Third, we require other words  where at least one of them must be in the article. Any of these words can be used in the articles which are  relevant to the testing of GUIs. It is important to note that all these words are connected by boolean ``\textbf{OR}'' making our search strategy more greedy, while all these words are also connected with \textbf{mobile} and \textbf{"gui testing"} by boolean ``\textbf{AND}'', thus targeting most relevant studies.

In Table~\ref{tbl:3}, we list the search results\footnote{searched in January, 2018} for the potentially relevant articles which are yielded by the specified search strategy.
 
\begin{table}[!hbpt]
\centering
\caption{\textbf{Search results for articles using generated search strategy.}
\label{tbl:3}}
\begin{tabular}{l|l|l} \hline
\textbf{Search source} & \textbf{Total sources searched} & \textbf{Total found articles} \\ \hline
Digital libraries & 11 & 958 \\ \hline
Search engines &  8 & 2,124 \\ \hline
Journals & 66 & 208 \\ \hline
Conferences & 34  & 349 \\  \hhline{=|=|=}
\multicolumn{1}{c|}{\textbf{Total}} & \textbf{119} & \textbf{3,639} \\ \hline
\end{tabular}
\end{table}

\subsection{Study selection}
\label{subsec:studysel}
\begin{figure}[!hbpt]
\centering
\includegraphics[keepaspectratio=true,scale=0.6,trim=60 40 50 30,clip=true]{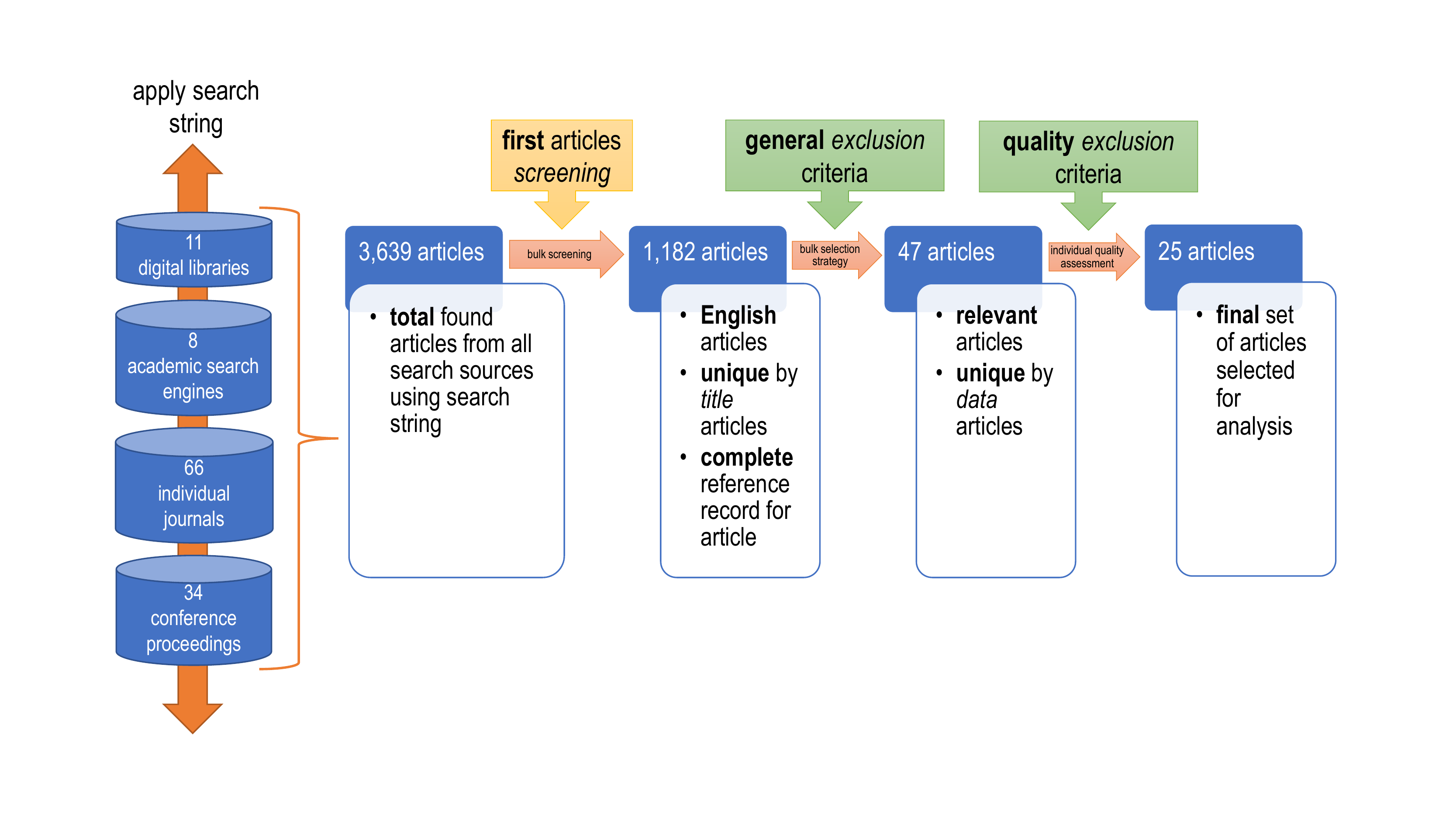}
\caption{Study selection: overall process of study selection including study search, screening, selection strategy, and quality assessment.}
\label{fig:1}
\end{figure}

In systematic reviews, once the potentially relevant primary studies have been identified, they further need to be assessed for their actual relevance to the topic of interest. For that purpose, study selection criteria are need to be developed to identify such primary studies which provide direct evidence of relevance about the research question(s) raised by the systematic review. Visually, we show the overall process of study selection in Figure~\ref{fig:1}.

\subsubsection{Selection strategy}
\label{subsubsec:selectionstr}
To select relevant primary studies, we first excluded all articles which have incomplete reference record information, non-English articles, and duplicates (i.e., articles with the same titles) from the found set of \R{3,639} articles. After filtering, we obtained \R{1,182} unique (by title) articles with complete reference record information, and all of which are written in English. Next, we identify relevant primary studies by reviewing their abstracts, semantics of titles and keywords. If the abstracts, titles, and keywords do not provide sufficient confidence, we also review conclusions of the target articles to conform their direct relevance to our topic of interest. From the set of \R{1,182} unique (by title) articles, we also excluded those which publish the same data. In particular, when there are two or more publications of the same data, we include the most complete one, and the others are excluded. So, after applying our general inclusion and exclusion criteria, we obtained \R{47} unique (by title and data) relevant articles. It is important to note that, from this systematic review, we excluded Patents, Books, Lecture Notes, Keynotes since they innately stay outside of interest for the systematic reviews due to their specific communication style which is not in line with a scientific research design. We also excluded Theses since their data has already been published either on conferences, or in journals.

In Table~\ref{tbl:4}, we provide a full list of general inclusion and exclusion criteria for primary studies selection.

\begin{table}[!hbpt]
\centering
\caption{\textbf{General  criteria for inclusion and exclusion of primary studies.}
\label{tbl:4}}
\begin{tabular}{l|l|l} \hline
\textbf{No.} & \textbf{Inclusion} & \textbf{Exclusion} \\ \hline
1 & Written in English & Other languages \\ \hline
2 & Technical\textcolor{blue}{*} paper & Other types of communications or works \\ \hline
3 & Conducted for mobile apps & Other applications \\ \hline
4 & Focus on automated\textcolor{blue}{**} functional GUI testing & Other types of GUI testing \\ \hline
5 & Provide complete\textcolor{blue}{***} relevant bibliography & Incomplete bibliography \\ \hline
\end{tabular}
\begin{tablenotes}
\footnotesize
  	     \item[] \hspace*{0.65cm}\textcolor{blue}{*} this includes full complete technical research papers. 
	     \item[] \hspace*{0.65cm}\textcolor{blue}{**} this includes studies with fully-automatic or semi-automatic testing techniques.
  	     \item[] \hspace*{0.65cm}\textcolor{blue}{***} this includes relevant references from 2007 year inclusively, onwards.
\end{tablenotes}
\end{table}

\subsubsection{Quality assessment}
\label{subsubsec:qualityassmnt}
In addition to the general inclusion and exclusion criteria, it is considered critical to assess the ``quality'' of the selected primary studies. The quality criteria provides more sophisticated details about inclusion and exclusion of the primary studies. However, there is no agreed definition of study ``quality'' that makes an initial difficulty for the quality assessment. Nevertheless, CRD Guidelines \R{\cite{khan2001undertaking}} and the Cochrane Reviewers' Handbook \R{\cite{cochrane2003}} both suggest that quality relates to the extent to which the primary study minimises bias, and maximises internal and external validity. As such, we prepare quality criteria which is aimed at assessing the extent to which primary studies have addressed their bias and validity. It is important to note that when we are  forming the quality criteria, we keep in mind that primary studies are often poorly report their results, so theoretically, it may not be possible to determine how to assess their quality, while  simply assuming that because something was not reported, it was not actually done. So, as suggested by Petticrew and Roberts \R{\cite{petticrew2008systematic}}, the quality criteria need to address not only the reporting quality, but also the methodological quality of the conducted research. We rigorously developed data quality questions to select most credible, well-designed, complete, and coherent research, so that we ensure that this systematic review analyses only primary studies with a reasonable quality.

Using our quality criteria, we assess quality of \R{47} relevant articles to confirm their inclusion to, or exclusion from this systematic review. Quality assessment helps to further evaluate the selected primary studies, and conclude to which extent they help to answer the RQs of this systematic review. Note that our quality criteria are used to assist primary studies selection by providing additional details for inclusion and exclusion criteria; we do not use the quality criteria to assist data analysis and synthesis. 

In Table~\ref{tbl:5}, we provide a full list of quality criteria for primary studies selection.

\begin{table}[!hbpt]
\centering
\caption{\textbf{Quality criteria for inclusion and exclusion of primary studies.}
\label{tbl:5}}
\begin{tabular}{l|l} \hline
\textbf{No.} & \textbf{Quality question}  \\ \hline
1& Is research problem clearly stated?  \\ \hline
2& Does it discover novel aspects not existing in other studies?  \\ \hline
3& Is research design properly documented?    \\ \hline
4& Is primary study outcome properly described and discussed?    \\ \hline
5& Have threats to study validity been discussed?\textcolor{blue}{*}    \\ \hline
\end{tabular}
\begin{tablenotes}
\footnotesize
	      \item[] \hspace*{3.05cm}\textcolor{blue}{*} at least internal and external threats.
\end{tablenotes}
\end{table}

To more accurately assess the data quality of the primary studies, we constructed a measurement scale for each quality question. For each data quality criterion, we assign weighting coefficient from [0...1] with step 0.1 depending on to which extent this particular study answers the quality question; `0' indicates no data quality, i.e., study does not answer this particular data quality question at all, while `1' indicates excellent data quality for this particular data quality question. For each data quality criterion, the weighting coefficients were independently assigned by 3 authors of this systematic review. Next, the assigned values were averaged to obtain the final weighting coefficient for each  quality criterion. By summing the obtained average values, we computed a total weight for each primary study. Upon discussion in our research group, for inclusion of a primary study, we set a minimum threshold for the total weight (quality index) of the primary study. If the total weight of primary study is \R{2.50 (i.e., 50\% of the possible maximum 5.00)} or more, we include the study in this systematic review, otherwise we consider that the study is of poor quality, and thus it should be excluded from consideration. If there have been any disagreements, they  were resolved during our group meeting with supervisor. 

Using our general and quality criteria for inclusion and exclusion of the primary studies, in total, we included \R{25} primary studies for analysis, while 22 primary studies were excluded. In Table~\ref{tbl:6}, we show the final numbers of included and excluded articles which have been obtained from  different publishing venues including Journals, Conferences, Workshops, Symposiums, Technical Reports, Magazines, While Papers, Internet, and Pre-prints. In Table~\ref{tbl:11}, we provide a full list of primary studies which are included in this systematic review for analysis. As shown in Table~\ref{tbl:11}, we describe each study by listing their attributes. While giving the primary studies descriptions, we group and sort them by year of publication  in descending order so that it can be easily seen a distribution of the articles by the publication year. Within a group, we sort the articles by their weight, and venue rank (for conferences) or impact factor (for journals) for the same weight, in descending order. In Table~\ref{tbl:15}, we provide a full list of the excluded articles with a rationale for exclusion. We group the articles by their type, and within each group, sort them  by venue rank (for conferences) or impact factor (for journals), and publication year for the same rank or impact factor, in descending order. The  articles have been excluded for various reasons, mainly because they have a different focus from this systematic review. However, being relevant to the topic of functional GUI testing in mobile, they  can be interesting to the reader. So, we maintain a list of the excluded relevant primary studies so that they can be found in ``References'' section of this systematic review.

\begin{table}[!hbpt]
\centering
\caption{\textbf{Final number of articles included in, and excluded from this systematic review.}
\label{tbl:6}}
\begin{tabular}{l|l|l} \hline
\textbf{Article type} & \textbf{Total included articles}& \textbf{Total excluded articles} \\ \hline
Journal &3  &5\\ \hline
Conference\textcolor{blue}{*} &20  &17\\  \hline
Other\textcolor{blue}{**} &2  &0\\ \hhline{=|=|=}
\multicolumn{1}{c|}{\textbf{Total}} & \textbf{\R{25}} &\textbf{\R{22}} \\ \hline
\end{tabular}
\begin{tablenotes}
\footnotesize
	      \item[] \hspace*{2.35cm}\textcolor{blue}{*} this includes Conferences,  Workshops, and Symposiums.
	      \item[] \hspace*{2.35cm}\textcolor{blue}{**} this includes Technical Reports, Magazines, While Papers, Internet, and Pre-prints.
\end{tablenotes}
\end{table}

{\fontsize{8}{10}\selectfont
\begin{landscape}
\begin{ThreePartTable}
\begin{TableNotes}
\footnotesize
	      \item[] \hspace*{0.05cm}\textcolor{blue}{*} these values indicate a total weight (quality index) of the article; it was obtained by summing the averaged weighting coefficients for each data quality criterion (see Table~\ref{tbl:5}).
	      \item[] \hspace*{0.05cm}\textcolor{blue}{**} in this column, we use \textit{Unknown} as we could not  identify publishing venue for the article.
	      \item[] \hspace*{0.05cm}\textcolor{blue}{***} for conferences, the rank values (A*, A, B, and C) are given in accordance with CORE2018 ``portal.core.edu.au/conf-ranks''; for journals, the impact factor (IF) values are given as of February, 2018.
\end{TableNotes}

\begin{longtable}{l|l|l|l|l|l}%
\caption{\textbf{List of primary studies included in this systematic review for analysis.}
\label{tbl:11}}\\ \hline
\multirow{2}{*}{}%
\textbf{Article} &
\textbf{Article} & 
\textbf{Article} &
\textbf{Article} & 
\textbf{Publishing} &
\textbf{Venue} \\ 
\textbf{Reference}&\textbf{Weight}\textcolor{blue}{*}&\textbf{Type}\textcolor{blue}{**}&\textbf{Year}&\textbf{Venue}&\textbf{Rank/IF}\textcolor{blue}{***}\\ \hline
\endfirsthead
\multicolumn{6}{c}%
{\tablename\ \thetable\ -- \textit{Continued from previous page}} \\ \hline
\multirow{2}{*}{}%
\textbf{Article} &
\textbf{Article} & 
\textbf{Article} &
\textbf{Article} & 
\textbf{Publishing} &
\textbf{Venue} \\ 
\textbf{Reference}&\textbf{Weight}\textcolor{blue}{*}&\textbf{Type}\textcolor{blue}{**}&\textbf{Year}&\textbf{Venue}&\textbf{Rank/IF}\textcolor{blue}{***}\\ \hline
\endhead
\hline 
\multicolumn{6}{r}{\textit{Continued on next page}} \\
\endfoot
\hline
\insertTableNotes
\endlastfoot

\rowcolor{mygray}\cite{autodroidcomb} &5.00& Journal & 2018   &Information and Software Technology & 2.694 \\ \hline
\rowcolor{mygray}\cite{androframe} &5.00& \textit{Unknown} & 2018 & Internet & -- \\ \hline

\cite{mobolic} &5.00& Journal & 2017 & Software: Practice and Experience   & 1.609 \\ \hline
\cite{stoat} &5.00& Conference & 2017  &Joint Meeting on Foundations of Software Engineering (ESEC/FSE)& A* \\ \hline
\cite{aimdroid} &5.00& Conference & 2017 & International Conference on Software Maintenance and Evolution (ICSME) & A \\ \hline
\cite{ehbdroid} &4.60& Conference & 2017 &International Conference on Automated Software Engineering (ASE) & A \\ \hline
\cite{patdroid} &4.00& Conference & 2017 &Joint Meeting on Foundations of Software Engineering (ESEC/FSE)  & A* \\ \hline
\cite{xdroid} &3.33&  Conference & 2017  &Computer Software and Applications Conference (COMPSAC)& B \\ \hline
\cite{land} &3.30& Conference & 2017  &International Conference on Software Quality, Reliability and Security (QRS)& B \\ \hline
\cite{droidwalker} &2.67& \textit{Unknown} & 2017 & arXiv preprint & -- \\ \hline

\rowcolor{mygray}\cite{sapienz} &5.00& Symposium & 2016 & International Symposium on Software Testing and Analysis (ISSTA) & A \\ \hline
\rowcolor{mygray}\cite{guicc} &4.00& Conference & 2016 & International Conference on Automated Software Engineering (ASE) & A \\ \hline 
\rowcolor{mygray}\cite{droiddev} &3.50& Conference & 2016 &  Asia--Pacific Software Engineering Conference (APSEC)& B \\ \hline
\rowcolor{mygray}\cite{mcrawlt} &3.27& Conference & 2016 & Software Engineering, Artificial Intelligence, Networking and Parallel/Distributed Computing (SNPD) & C \\ \hline
\rowcolor{mygray}\cite{gat} &2.77& Conference & 2016   &Asia--Pacific Software Engineering Conference (APSEC)& B \\ \hline 
 
\cite{mobiguitar} &3.57& Journal & 2015 & IEEE Software & 2.190 \\ \hline
\cite{sigdroid} &3.00& Symposium & 2015 &International Symposium on Software Reliability Engineering (ISSRE) &A \\ \hline
 
\rowcolor{mygray}\cite{evodroid} &4.00& Symposium & 2014   &International Symposium on Foundations of Software Engineering (FSE)& A* \\ \hline
\rowcolor{mygray}\cite{adautomation} &3.73& Conference & 2014 & International Conference on Software Security and Reliability (SERE) & B \\ \hline

\cite{swifthand} &4.27& Conference & 2013 & International Conference on Object-Oriented Programming, Systems, Languages, and Applications (OOPSLA) & A* \\ \hline
\cite{collider} &4.17& Symposium & 2013 & International Symposium on Software Testing and Analysis  (ISSTA) & A \\ \hline
\cite{a3e} &4.00& Conference & 2013 &  International Conference on Object-Oriented Programming, Systems, Languages, and Applications (OOPSLA) & A* \\ \hline 
\cite{orbit} &4.00& Conference & 2013 & Fundamental Approaches to Software Engineering (FASE) & B \\ \hline
\cite{dynodroid} &3.80& Conference & 2013 & Joint Meeting on Foundations of Software Engineering (ESEC/FSE) & A* \\ \hline
 
\rowcolor{mygray}\cite{acteve} &4.87& Symposium & 2012 & International Symposium on the Foundations of Software Engineering (FSE) & A* \\ \hline
\end{longtable}
\end{ThreePartTable}
\end{landscape}
}

{\fontsize{8}{10}\selectfont
\begin{landscape}
\begin{ThreePartTable}
\begin{TableNotes}
\footnotesize
	      \item[] \hspace*{0.05cm}\textcolor{blue}{*} in this column, we use \textit{Unknown} as we could not  identify publishing venue for the article.
	      \item[] \hspace*{0.05cm}\textcolor{blue}{**} for conferences, the rank values (A*, A, B, and C) are given in accordance with CORE2018 ``portal.core.edu.au/conf-ranks''; for journals, the impact factor (IF) values are given as of February, 2018.
\end{TableNotes}

\begin{longtable}{l|l|l|l|l|l}%
\caption{\textbf{List of primary studies excluded from this systematic review.}
\label{tbl:15}}\\ \hline
\multirow{2}{*}{}%
\textbf{Article} & 
\textbf{Article} &
\textbf{Article} & 
\textbf{Publishing} &
\textbf{Venue} &
\textbf{Rationale} \\ 
\textbf{Reference}&\textbf{Type}\textcolor{blue}{*}&\textbf{Year}&\textbf{Venue}&\textbf{Rank/IF}\textcolor{blue}{**}&\textbf{for Exclusion}\\ \hline
\endfirsthead
\multicolumn{6}{c}%
{\tablename\ \thetable\ -- \textit{Continued from previous page}} \\ \hline
\multirow{2}{*}{}%
\textbf{Article} & 
\textbf{Article} &
\textbf{Article} & 
\textbf{Publishing} &
\textbf{Venue} \\ 
\textbf{Reference}&\textbf{Type}\textcolor{blue}{*}&\textbf{Year}&\textbf{Venue}&\textbf{Rank/IF}\textcolor{blue}{**}\\ \hline
\endhead
\hline 
\multicolumn{6}{r}{\textit{Continued on next page}} \\
\endfoot
\hline
\insertTableNotes
\endlastfoot

\rowcolor{mygray}\cite{excludeMUTATION} & Journal & 2014 & Information and Software Technology &2.694 &  mutation testing, non-functional \\ \hline

\rowcolor{mygray}\cite{mobolic15} & Journal & 2017 & Journal of Systems and Software & 2.444 &  non-technical research, comparison framework \\ \hline

\rowcolor{mygray}\cite{excludeMANUAL} & Journal & 2014 & Journal of Systems and Software & 2.444 &  manual testing \\ \hline

\rowcolor{mygray}\cite{excludeVOG} & Journal & 2017 & IEEE Software & 2.190 & predominant manual,  capture-replay approach \\ \hline

\rowcolor{mygray}\cite{excludeIMPACT} & Journal & 2017 &  Software Quality Journal & 1.816 & pattern-based testing,   non-functional \\ \hline

\cite{word2vec} & Conference & 2017 & International Conference on Software Engineering (ICSE) & A* &  textual input generation, non-functional \\ \hline

\cite{excludeMOBIPLAY} & Conference & 2016 &  International Conference on Software Engineering  (ICSE)& A* & predominant manual,  capture-replay approach \\ \hline

\cite{excludeRERAN} & Conference & 2013 & International Conference on Software Engineering (ICSE) &A*  & predominant manual, capture-replay approach  \\ \hline

\cite{sketch} & Conference & 2017 &  International Conference on Automated Software Engineering  (ASE)& A & new idea, short communication \\ \hline

\cite{excludeAPPCHECK} & Conference & 2017 &  International Conference on Web Services (ICWS)& A & predominant manual, capture-replay approach \\ \hline

\cite{excludeTESTMINER} & Conference & 2017 &  International Conference on Automated Software Engineering  (ASE)& A & new idea, short communication \\ \hline

\cite{excludeATOM} & Conference & 2017 & International Conference on Software Testing, Verification and Validation (ICST) &A  & regression testing, non-functional \\ \hline

\cite{androframe7} & Conference & 2016 & International Conference on Software Testing, Verification and Validation (ICST) &A  & non-functional \\ \hline

\cite{excludeMONKEYLAB} & Conference & 2015 & Working Conference on Mining Software Repositories (MSR) &A  & predominant manual, capture-replay approach  \\ \hline

\cite{quantum} & Conference & 2014 &  International Conference on Software Testing, Verification and Validation (ICST)& A & oracle generation, non-functional \\ \hline

\cite{excludeTEMA} & Conference & 2011 & International Conference on Software Testing, Verification and Validation (ICST) &A & non-technical research, industrial case study \\ \hline

\cite{amoga} & Conference & 2016 & International Conference on Advances in Mobile Computing and Multimedia (MoMM) & B & unsatisfactory article quality \\ \hline

\cite{cadage} & Conference & 2015 &  Computer Software and Applications Conference (COMPSAC)& B & unsatisfactory article quality \\ \hline

\cite{excludeUGA} & Conference & 2014 & Asia-Pacific Software Engineering Conference (APSEC) &B  & predominant manual, capture-replay approach  \\ \hline

\cite{puma} & Conference & 2014 & International Conference on Mobile Systems, Applications, and Services (MobiSys) &B  & predominant manual, user-programmable framework  \\ \hline

\cite{excludeDROIDBOT} & Conference & 2017 & International Conference on  Software Engineering Companion (ICSE-C) &--  & compatibility testing, non-functional  \\ \hline

\cite{excludeDROIDMATE} & Conference & 2016 & International Conference on  Mobile Software Engineering and Systems (MOBILESoft) &--  & non-functional \\ \hline
\end{longtable}
\end{ThreePartTable}
\end{landscape}
}

\subsection{Data extraction}
\label{subsec:dataextr}
\begin{figure}[!hbpt]
\centering
\includegraphics[keepaspectratio=true,scale=0.6,trim=100 100 160 0,clip=true]{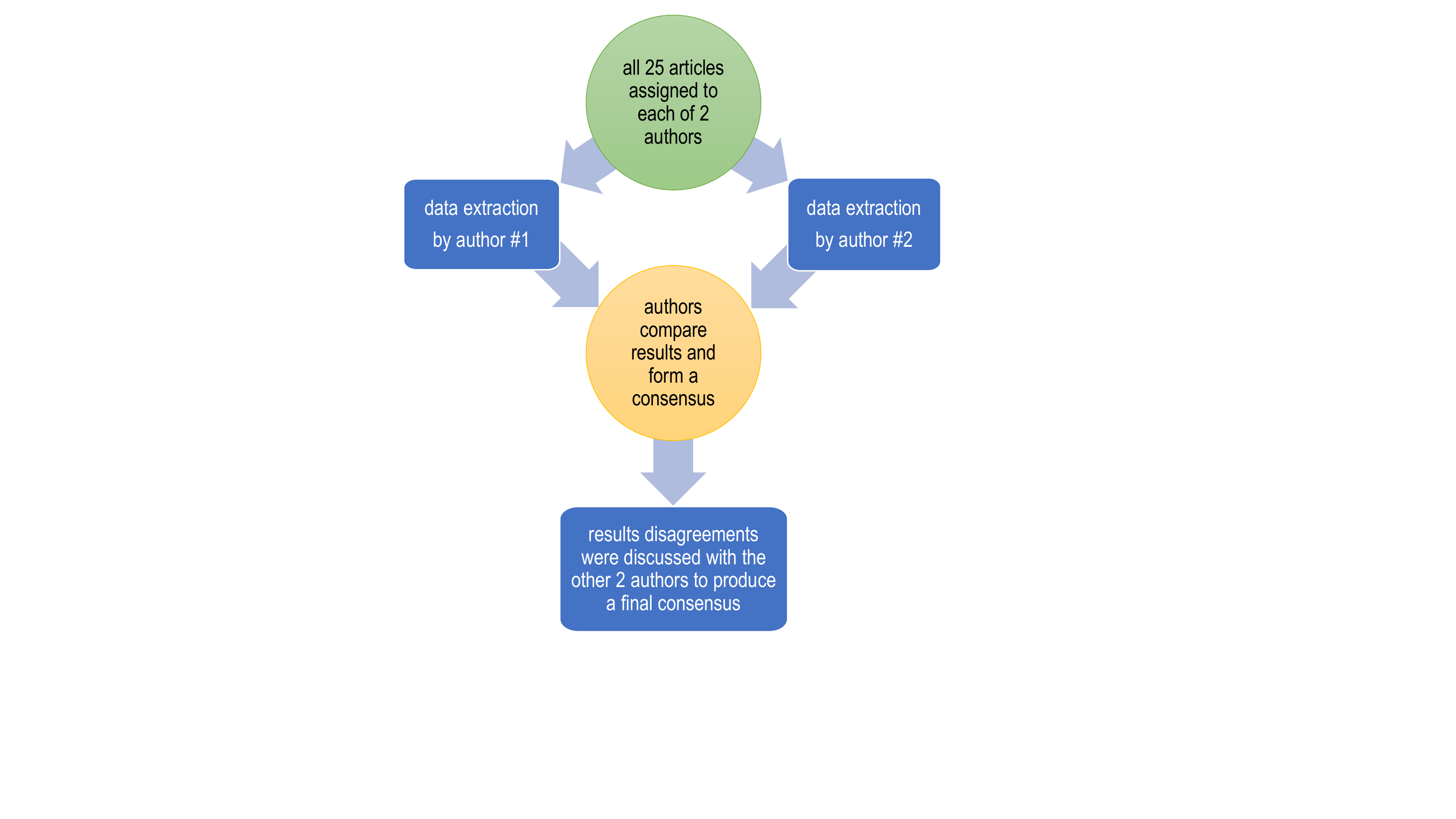}
\caption{Data extraction: overall process of data extraction, and consensus formation.}
\label{fig:2}
\end{figure}

The objective of this stage is to collect all the information needed to address the RQs. Tabulating the extracted data is a useful instrument of aggregation, so we design data extraction forms (tables) to accurately record the information extracted from  the primary studies. Prior to forming final data extraction forms, we piloted them on a sample of primary studies. The pilot studies help to assess the completeness, clarity and structure of the data forms \cite{slrguide}. For the data extraction, we assigned 2 authors. Each author extracted data independently from all \R{25} selected primary studies. Next, the filled data forms were collected from the authors, and the provided data was compared. If there have been any conceptional mismatches, they were resolved during our discussion with the other 2 authors of this systematic review. Visually, we show the data extraction process and consensus formation in Figure~\ref{fig:2}.

We structured our tables in such a way helping to highlight similarities and differences between primary studies outcomes in one place. In particular, for every RQ of our review, we prepared a separate table which includes the data relevant to each RQ. Below, we show examples of data extraction forms which include headers of the respective tables, Table~\ref{tbl:8}, Table~\ref{tbl:9}, and Table~\ref{tbl:10}. We explain a functional meaning each of the columns in the tables, based on the definitions provided in \cite{utting}. First, we describe columns which are common for all the tables. And next, we describe columns which are specific to each individual table.

Here, we describe common columns for all the tables. 
\begin{itemize}
\item[] ``\textbf{Article Reference}''. It shows a reference number of the primary study which is assigned in this systematic review. So the primary study can be found in section ``References'' of this systematic review.

\item[] ``\textbf{Artefact}''. It shows an acronym of the  technique, tool, approach or method which is discussed in the primary study.
\end{itemize}

Here, we describe columns which are specific to Table~\ref{tbl:8}. 
\begin{itemize}
\item[] ``\textbf{Model Paradigm}''. It describes nature of the built model. It shows which modelling notations are used to describe the behaviour of the target apps for test generation purposes.

\item[] ``\textbf{Test Generation Approach}''. It shows how tests are derived from a built model. It may also combine several approaches to facilitate the non-trivial task of automated test generation from a model.

\item[] ``\textbf{Test Generation Criteria}''. It defines test criteria which are used to control the generation of tests.  These criteria indirectly define properties of the generated test suites, including their fault detection capability, cardinality, and structural complexity.

\item[] ``\textbf{Textual Input Generation}''. It shows which kinds of textual user-inputs can be generated by the artefact. The input generation process itself can be automated or manual, while the input kinds can be random, concrete, predefined, contextual, or others. By default, all the generated inputs are ``Automated''. However, if an artefact requires human intervention during the testing process, we indicate it as ``Manual''.

\item[] ``\textbf{Code Coverage Metric}''. It shows which code coverage metric is used in the primary study. In this systematic review, we extract data only for ``basic-block'', and ``line (statement)'' metrics. If a primary study does not provide such metrics, we indicate ``\textit{NA}''. In this systematic review we use  ``basic-block'', and ``line (statement)'' metrics because they are de facto fundamental  means of the  effectiveness assessment. The other code coverage metrics such as ``class'', ``method'', ``branch'', or model coverage metrics such as ``activity'', ``transitions'', ``states'',  ``events'', or  ``sequences of events'', as excluded since they innately cannot give an adequate assessment of the artefact effectiveness.

\item[] ``\textbf{Code Coverage (average value), \%}''. It shows an average value for the benchmark apps used in a primary study. We computed the average values for the apps which are reported in the primary studies with ``basic-block'', and ``line (statement)'' code coverage metrics. If a primary study does not provide values for such metrics, we indicate ``\textit{NA}''.

\item[] ``\textbf{Effectiveness (relative estimation)}''. It shows a relative estimation of effectiveness for the artefact. The relative estimation of effectiveness is given based on the average code coverage values which are extracted from the data provided in primary studies. More details about the effectiveness estimation we provide further in this section.
\end{itemize}

Here,  we describe columns which are specific to Table~\ref{tbl:9}. 
\begin{itemize}
\item[] ``\textbf{Testing Approach}''. It shows which model is used for the automated testing. We identify 2 main models: one is derived from the ``GUI'' (user interface flow), and another one is derived from the ``Code'' (source or binary) of the target app.

\item[] ``\textbf{Testing Technique}''. It shows in which manner an artefact performs an exploration of the built model. We identify 2 main approaches: one is ``Systematic'', where the artefact implements guided exploration, and another one is ``Random'', where the artefact implements random-based exploration strategy of the built model.

\item[] ``\textbf{Search Algorithm}''. It shows which algorithm is used, or based on to guide the exploration process. We identify 2 main algorithms: one is ``Guided'' which uses various model heuristics to guide the search, and another one is ``Random'' which is based on uniform random and probabilistic events generation.

\item[] ``\textbf{Termination Condition}''. It shows a condition which is to be satisfied to terminate the testing process. The termination condition can be determined either by the model properties, or  manually by the user.

\item[] ``\textbf{Benchmark Size}''. It shows a total number of apps for which the respective data was extracted from a primary study.

\item[] ``\textbf{Execution Time (per app), mins}''. It shows minimum, maximum, exact or average time taken for each app from the benchmark used in a primary study. If a primary study does not provide execution time values, we indicate ``\textit{NA}'', or we indicate  ``\textit{N/A}'' if the execution time is not applicable to the artefact.

\item[] ``\textbf{Efficiency (relative estimation)}''.  It shows a relative estimation of efficiency of the artefact. The relative estimation of efficiency is given based on the execution time values which are extracted from the data provided in primary studies. More details about the efficiency estimation we provide further in this section.
\end{itemize}

In Table~\ref{tbl:10}, we collate all the numerical data which is relevant to practicality. This table  facilitates RQ\#3 by giving a visual analysis of the derived data from Table~\ref{tbl:8}, and Table~\ref{tbl:9}. So we do not repeat descriptions for the matching columns from Table~\ref{tbl:8}, and Table~\ref{tbl:9}, and only describe unique columns which are specific to Table~\ref{tbl:10}.
\begin{itemize}
\item[] ``\textbf{Practicality (relative estimation)}''. It shows a relative estimation of practicality of the artefact. The relative estimation of practicality is given based on the averaged effectiveness and efficiency which are extracted from Table~\ref{tbl:8}, and Table~\ref{tbl:9}, respectively. More details about the practicality estimation we provide further in this section. 
\end{itemize}

\begin{landscape}
\begin{table}[!hbpt]
\centering
\caption{\textbf{Data extraction form for RQ\#1 and its secondary questions: Effectiveness estimation.}
\label{tbl:8}}
\begin{tabular}{l|l|l|l|l|l|l|l|l} \hline
\multirow{2}{*}{}%
\textbf{Article} &
\textbf{Artefact} &
\textbf{Model} &\textbf{Test Generation} &\textbf{Test Generation}&\textbf{Textual Input} &\textbf{Code Coverage} &\textbf{Code Coverage} &\textbf{Effectiveness} \\ 
\textbf{Reference}&&\textbf{Paradigm}&\textbf{Approach} &\textbf{Criteria}&\textbf{Generation} &\textbf{Metric} &\textbf{(average value), \%} &\textbf{(relative estimation)} \\
\hline

$\cdots$  &$\cdots$  & $\cdots$ &$\cdots$  & $\cdots$ & 
$\cdots$ &$\cdots$ & $\cdots$ & $\cdots$ \\ \hline
\end{tabular}
\end{table}

\begin{table}[!hbpt]
\centering
\caption{\textbf{Data extraction form for RQ\#2 and its secondary questions: Efficiency estimation.}
\label{tbl:9}}
\begin{tabular}{l|l|l|l|l|l|l|l|l} \hline
\multirow{2}{*}{}%
\textbf{Article} &
\textbf{Artefact} &
\textbf{Testing} & \textbf{Testing} & \textbf{Search} & \textbf{Termination} &\textbf{Benchmark} & \textbf{Execution Time} & \textbf{Efficiency} \\ 
\textbf{Reference}&&\textbf{Approach} & \textbf{Technique} & \textbf{Algorithm} & \textbf{Condition} &\textbf{Size} & \textbf{(per app), mins} & \textbf{(relative estimation)} \\ \hline

$\cdots$ & $\cdots$  & $\cdots$ &$\cdots$  & $\cdots$ & 
$\cdots$ &$\cdots$ & $\cdots$ & $\cdots$ \\ \hline

\end{tabular}
\end{table}

\begin{table}[!hbpt]
\centering
\caption{\textbf{Data extraction form for RQ\#3 and its secondary questions: Practicality estimation.}
\label{tbl:10}}
\begin{tabular}{l|l|l|l|l|l|l} \hline
\multirow{2}{*}{}%
\textbf{Article} &
\textbf{Artefact} &
\textbf{Code Coverage}\textcolor{blue}{*} & \textbf{Execution Time}\textcolor{blue}{**} & \textbf{Effectiveness}\textcolor{blue}{*} &\textbf{Efficiency}\textcolor{blue}{**}  & \textbf{Practicality} \\ 
\textbf{reference}&&\textbf{(average value), \%} & \textbf{(per app), mins} & \textbf{(relative estimation)}& \textbf{(relative estimation)} & \textbf{(relative estimation)} \\ 
\hline

$\cdots$ &$\cdots$ & $\cdots$  & $\cdots$ &$\cdots$  & $\cdots$ & $\cdots$ \\ \hline

\end{tabular}
\begin{tablenotes}
\footnotesize
	      \item[] \hspace*{1.05cm}\textcolor{blue}{*} values for these columns will be taken from the respective columns in Table~\ref{tbl:8}.
	      \item[] \hspace*{1.05cm}\textcolor{blue}{**}  values for these columns will be taken from the  respective columns in Table~\ref{tbl:9}.
\end{tablenotes}
\end{table}
\end{landscape}

In this systematic review, to answer our RQs, we provide a relative estimation of effectiveness, efficiency, and practicality of the automated mobile testing techniques. In fact, there is no clear way how to identify absolute values for effectiveness, efficiency, and practicality. As such, the estimation we give is relative because it is solely based on the data extracted from the primary studies. We estimate effectiveness, efficiency, and practicality by comparing their relevant extracted (or deduced) data for each of the artefacts against a specified range of values (intervals). How the intervals have been decided, we explain below. We also estimate practicality each of the artefacts by averaging the relative values of effectiveness, and efficiency.

As we can see, authors of different primary studies evaluate their techniques using their own benchmark apps, and experimental environments. For example, different primary studies use apps with different code size (lines of code), GUIs and code complexity. Also, different primary studies use different execution  environments (e.g., physical mobile device, or mobile emulator), computational resources (e.g., desktop, server, or cloud), and certain techniques may require humans to participate. So, all these variations make it difficult to uniformly give an estimation of the effectiveness, efficiency, and practicality. To find a solution, we make a valid assumption which is based on the fact that the authors of their primary studies choose such benchmark apps, and set-up such experimental environments which are most suited for the purpose of demonstrating an effectiveness, efficiency, and practicality of their proposed testing techniques. As such, every testing technique is expected to show its best performance results. Based on this fact, we believe that our assumption is reasonable and valid, and thus our relative estimation should reflect a true performance of the testing techniques. However, it is important to note that the extracted numerical data values for the effectiveness, efficiency, and practicality estimation are only representatives of the techniques performance. As such, they should not be considered as absolute performance values which are persistent across different apps, and experimental environments.
    
\paragraph{Effectiveness}
Effectiveness  is one of the critical characteristics of the model-based testing techniques. It is usually determined by the inferred model, quality of which must be persistent across different testing apps. Based on this fact, we assume that our relative estimation of effectiveness is persistent as well, and must not vary depending on the apps. For example, using high quality models, the effectiveness must not be affected by the apps code size (lines of code), GUIs and code complexity. However, practice shows that there could be a chance when specific apps may affect effectiveness (usually to lower side) due to possible incompleteness of the inferred model. 

Based on our practical experience and observations, we conditionally determine the code coverage intervals to   estimate an effectiveness of the testing techniques as follows.
\begin{itemize}
\item[] \CCCCC ~|~ very high (\textgreater95\%) 
\item[] \CCCC ~|~ high (86\%--95\%)
\item[] \CCC ~|~ medium (71\%--85\%)
\item[] \CC ~|~ low (51\%--70\%)
\item[] \C ~|~ very low (\textless50\%)
\end{itemize}

It is important to note that these intervals are only valid for the code coverage which is measured in lines (statements), or basic-blocks. In fact, lines (statements) and basic-blocks are closely related to each other code coverage metrics. In particular, one line of code may correspond to several basic-blocks, and otherwise, one basic-block may be included in multiple lines. So, based on this fact and our observations, we found that absolute difference in code coverage between lines and basic-blocks is minimal (i.e., its absolute value varies around $\pm$5\%), so that the defined intervals can be used for both metrics. However, it is important to note that none of other code coverage metrics should be used on these intervals. 

\paragraph{Efficiency} 
Efficiency is another critical characteristic of the model-based testing techniques. In practice, it is hard to estimate efficiency because it depends on various factors which are usually not persistent. For example, similar to the effectiveness, efficiency also may vary depending on the  testing apps code size, their GUI and code complexity, experimental environments. In addition, efficiency may depend on human factors such as user's programming skills and knowledge,  computational complexity of the implemented technique, search algorithm optimization, system design, and others.

Based on our practical experience and observations, we conditionally determine the execution time intervals to   estimate an efficiency of the testing techniques as follows.
\begin{itemize}
\item[] \ccccc ~|~ very high (\textless5 mins)
\item[] \cccc ~|~ high (5--10 mins)
\item[] \ccc ~|~ medium (11--20 mins)
\item[] \cc ~|~ low (21--35 mins)
\item[] \ca ~|~ very low (\textgreater35 mins)
\end{itemize}
It is important to note that these intervals are only valid for automated testing techniques. The execution time of manual techniques should not be evaluated on these intervals.

\paragraph{Practicality} 
Practicality is another critical characteristic of the model-based testing techniques. Generally, practicality depends on the effectiveness and efficiency of the testing techniques. In this systematic review, we consider importance of effectiveness and efficiency to be equivalent. So, we compute practicality relative value as an average of effectiveness and efficiency to give a sense of possible practicality of the testing techniques. 

Based on the estimated effectiveness and efficiency, we give a relative estimation of the practicality for each testing technique. For that purpose, we determine 5 base levels to estimate a practicality of the testing techniques as follows.
\begin{itemize}
\item[] \clllll ~|~ very high (likely to be practical)
\item[] \cllll ~|~ high (may be practical)
\item[] \clll ~|~ medium (could be practical)
\item[] \cll ~|~ low (may not be practical)
\item[] \cl ~|~ very low (unlikely to be practical)
\end{itemize}
It is important to note that these base levels are only valid for automated testing techniques. The practicality of manual techniques should not be evaluated on these levels.

\subsection{Data synthesis}
\label{subsec:datasynth}
Data synthesis is a process which involves collating and summarising the results of the included primary studies. In our systematic review, since our selected primary studies are qualitative  (i.e., descriptive in nature \cite{BRERETON2007571}), we describe their natural language results, numerical, and conclusions. In particular, we use ``Line of argument synthesis'' approach \cite{noblit1988meta}, where we first individually analyse each primary study, and next analyse the primary studies as a whole. For that purpose, we fill the tables, Table~\ref{tbl:12}, Table~\ref{tbl:13}, and Table~\ref{tbl:14}, with the respective to the RQs data which is extracted from the selected primary studies. For a better visual representation of the large tables, in Table~\ref{tbl:12}, we group the artefacts by ``\textbf{Model Paradigm}'' and sort each group by values in ``\textbf{Code Coverage (average value), \%}'' (from highest to lowest); in Table~\ref{tbl:13}, we group the artefacts by ``\textbf{Testing Approach}'' and sort each group by values in ``\textbf{Execution Time (per app), mins}'' (from shortest to longest); in Table~\ref{tbl:14}, we sort and group the artefacts by ``\textbf{Practicality (relative estimation)}'' (from highest to lowest).

By filling each of the tables, we analyse each primary study. We tabulated the extracted data in the manner helping to collate, and summarize the results of the primary studies in order to answer our secondary RQs. The answers to the primary RQs are based on the answers to the secondary RQs. However, answers to the primary RQs cannot be directly found in the primary studies, so the answers to the primary RQs are to be deduced by the authors of this systematic review. Based on the extracted data, we answer our primary RQs in ``Results'' section by analysing the primary studies as a whole.

{\fontsize{8}{10}\selectfont
\begin{landscape}
\begin{ThreePartTable}
\begin{TableNotes}
\footnotesize
	      \item[] \hspace*{0.05cm}\textcolor{blue}{*} it is based on the low-level atomic genes, and high-level motif genes.
\end{TableNotes}

\setlength\LTleft{-15pt}
\begin{longtable}{l|l|l|l|l|l|l|l|l}%
\caption{\textbf{Extracted data for RQ\#1 and its secondary questions: Effectiveness estimation.}  
\label{tbl:12}} \\ \hline
\multirow{2}{*}{}%
\textbf{Article} &
\textbf{Artefact} &
\textbf{Model} &\textbf{Test Generation} &\textbf{Test Generation}&\textbf{Textual Input} &\textbf{Code Coverage} &\textbf{Code Coverage} &\textbf{Effectiveness} \\ 
\textbf{Reference}&&\textbf{Paradigm}&\textbf{Approach} &\textbf{Criteria}&\textbf{Generation} &\textbf{Metric} &\textbf{(average value), \%} &\textbf{(relative estimation)} \\ \hline
\endfirsthead
\multicolumn{9}{c}%
{\tablename\ \thetable\ -- \textit{Continued from previous page}} \\ \hline
\multirow{2}{*}{}%
\textbf{Article} &
\textbf{Artefact} &
\textbf{Model} &\textbf{Test Generation} &\textbf{Test Generation}&\textbf{Textual Input} &\textbf{Code Coverage} &\textbf{Code Coverage} &\textbf{Effectiveness} \\ 
\textbf{Reference}&&\textbf{Paradigm}&\textbf{Approach} &\textbf{Criteria}&\textbf{Generation} &\textbf{Metric} &\textbf{(average value), \%} &\textbf{(relative estimation)} \\ \hline
\endhead
\hline 
\multicolumn{9}{r}{\textit{Continued on next page}} \\
\endfoot
\hline
\insertTableNotes
\endlastfoot

\rowcolor{mygray}\multirow{1}{*}{}
\cite{dynodroid} & 
Dynodroid  & 
Random--based  & 
Random generation & Fault detection &Random,& Line (statement) & 55 & \CC \\ \rowcolor{mygray} 
&&&(feedback--directed)&&Manual&&&\\ \hline

\rowcolor{mygray}\multirow{1}{*}{}
\cite{autodroidcomb} & 
Autodroid   & 
Random--based  & 
Random generation & Fault detection &Random,& Basic-block & 52 & \CC \\  \rowcolor{mygray}
&(Frequency)&&&&Predefined&&&\\ \hline

\rowcolor{mygray}\multirow{1}{*}{}
\cite{xdroid} & 
Xdroid  & 
Random--based  & 
Random generation & Fault detection &Random,& Line (statement) & 39 & \C \\  \rowcolor{mygray}
&&&&&Manual&&&\\ \hline

\multirow{1}{*}{}
\cite{stoat} & 
Stoat  & 
Stochastic  & 
Random generation & Fault detection &Random& Line (statement) & 61 & \CC \\  
&&&(based on Gibbs sampling)&&&&&\\ \hline

\multirow{1}{*}{}
\cite{aimdroid} & 
AimDroid  & 
Stochastic  & 
Random generation & Fault detection &Random& \textit{NA} & \textit{NA} & -- \\  
&&&(based on reinforcement&&&&&\\ 
&&& learning)&&&&&\\ \hline

\multirow{1}{*}{}
\cite{androframe} & 
AndroFrame  & 
Stochastic  & 
Random generation  & Fault detection &Random& \textit{NA} & \textit{NA} & -- \\  
&&&(based on Q--Matrix)&&&&&\\ \hline

\rowcolor{mygray}\multirow{1}{*}{}
\cite{mobiguitar} & 
MobiGUITAR  & 
State--based  & 
Model--checking & Structural model  &Random, & \textit{NA} & \textit{NA} & -- \\  \rowcolor{mygray}
&&&&coverage&User--predefined&&&\\ \hline

\multirow{2}{*}{}
\cite{mobolic} & 
Mobolic  & 
Transition--based  & 
Model--checking and & Structural model &Random,& Basic--block & 92 & \CCCC \\
&&&constraint solving&coverage&Concrete,&&& \\ 
&&&&&UI-context-aware,&&&\\ 
&&&&&User-predefined&&&\\ \hline

\multirow{1}{*}{}
\cite{droiddev} & 
DroidDEV & 
Transition--based  & 
Model--checking & Structural model &Random,& Basic--block & 91 & \CCCC \\ 
&&&&coverage&UI-context-aware,&&&\\ 
&&&&&User-predefined&&&\\ \hline

\multirow{1}{*}{}
\cite{orbit} & 
ORBIT  & 
Transition--based  & 
Model--checking & Structural model  &Random& Line (statement) & 78 & \CCC \\
&&&&coverage&&&&\\ \hline

\multirow{1}{*}{}
\cite{mcrawlt} & 
MCrawlT  & 
Transition--based  & 
Model--checking & Structural model  &Random& Line (statement) & 65 & \CC \\  
&&&&coverage&&&&\\ \hline

\multirow{1}{*}{}
\cite{land} & 
LAND  & 
Transition--based  & 
Model--checking & Structural model  &Random& Line (statement) & 58 & \CC \\*  
&&&&coverage&&&&\\ \hline

\multirow{1}{*}{}
\cite{droidwalker} & 
DroidWalker  & 
Transition--based  & 
Model--checking & Structural model  &Random& Line (statement) & 57 & \CC \\  
&&&&coverage&&&&\\ \hline

\multirow{1}{*}{}
\cite{autodroidcomb} & 
Autodroid   & 
Transition--based  & 
Model--checking & Structural model &Random,& Basic-block & 57 & \CC \\  
&(Combinatorial)&&&coverage&Predefined&&&\\ \hline

\multirow{1}{*}{}
\cite{guicc} & 
GUICC  & 
Transition--based  & 
Model--checking & Structural model  &Random& Line (statement) & 43 & \C \\  
&&&&coverage&&&&\\ \hline

\multirow{1}{*}{}
\cite{swifthand} & 
SwiftHand  & 
Transition--based  & 
Model--checking & Structural model  &Random, & \textit{NA} & \textit{NA} & -- \\  
&&&&coverage&Predefined&&&\\ \hline

\multirow{2}{*}{}
\cite{a3e} & 
A\textsuperscript{3}E  & 
Transition--based  & 
Model--checking & Structural model  &Random& \textit{NA} & \textit{NA} & -- \\
&(Depth--first)&&&coverage&&& \\  \hline

\multirow{1}{*}{}
\cite{adautomation} & 
ADAutomation  & 
Transition--based  & 
Model--checking & Structural model  &Random& \textit{NA} & \textit{NA} & -- \\  
&&&&coverage&&&&\\ \hline

\multirow{1}{*}{}
\cite{gat} & 
GAT  & 
Transition--based  & 
Model--checking & Structural model  &Random& \textit{NA} & \textit{NA} & -- \\  
&&&&coverage&&&&\\ \hline

\rowcolor{mygray}\multirow{1}{*}{}
\cite{sapienz} & 
Sapienz  & 
Genetic--based\textcolor{blue}{*}  & 
Multi--objective & Fault detection  &Random& Line (statement) & 53 & \CC \\  \rowcolor{mygray}
&&&search--based&&&&&\\ \rowcolor{mygray}
&&&(Pareto--optimal)&&&&&\\ \hline

\multirow{1}{*}{}
\cite{evodroid} & 
EvoDroid  & 
Control--flow  & 
Search--based algorithms & Structural code  &Random& Line (statement) & 81 & \CCC  \\  
&&(call graph--based)&&coverage&&&&\\ \hline

\multirow{2}{*}{}
\cite{sigdroid} & 
SIG--Droid  & 
Control--flow & 
Symbolic execution and & Structural code &Random,& Line (statement) & 78 & \CCC \\
&&(call graph--based)&constraint solving&coverage&Concrete&& \\ \hline

\multirow{1}{*}{}
\cite{patdroid} & 
PATDroid  & 
Control-- and   & 
Model--checking & Structural code  &Random& Line (statement) & 60 & \CC \\ 
&&data--flow&&coverage&&&&\\ 
&&(permission--aware)&&&&&&\\ \hline

\multirow{1}{*}{} 
\cite{ehbdroid} & 
EHBDroid  & 
Control--flow   & 
Model--checking & Structural code  &Random& Line (statement) & 57 & \CC  \\ 
&&(event handler--based)&&coverage&&&&\\ \hline

\multirow{2}{*}{}
\cite{a3e} & 
A\textsuperscript{3}E  & 
Control-- and  & 
Model--checking & Structural model  &Random& \textit{NA} & \textit{NA} & -- \\
&(Targeted)&data--flow&&coverage&&& \\  \hline

\multirow{2}{*}{}
\cite{acteve} & 
ACTEve  & 
Control-- and  & 
Symbolic execution and & Structural code  &Random,& \textit{NA} & \textit{NA} & -- \\  
&&data--flow&constraint solving&coverage&Concrete&&\\ \hline

\multirow{1}{*}{}
\cite{collider} & 
Collider  & 
Control--flow  &
Symbolic execution  and & Structural code  &Random,& \textit{NA} & \textit{NA} & -- \\  
&&(call graph--based)&constraint solving&coverage&Concrete&&\\ \hline

\end{longtable}
\end{ThreePartTable}
\end{landscape}
}

{\fontsize{8}{10}\selectfont
\begin{landscape}
\begin{ThreePartTable}
\begin{TableNotes}
\footnotesize
	      \item[] \hspace*{0.05cm}\textcolor{blue}{*} it combines random and systematic exploration strategies.
\end{TableNotes}

\setlength\LTleft{-10pt}
\begin{longtable}{l|l|l|l|l|l|l|l|l}%
\caption{\textbf{Extracted data for RQ\#2 and its secondary questions: Efficiency estimation.}
\label{tbl:13}} \\ \hline
\multirow{2}{*}{}%
\textbf{Article} &
\textbf{Artefact} &
\textbf{Testing} & \textbf{Testing} & \textbf{Search} & \textbf{Termination} &\textbf{Benchmark} & \textbf{Execution Time} & \textbf{Efficiency} \\ 
\textbf{Reference}&&\textbf{Approach} & \textbf{Technique} & \textbf{Algorithm} & \textbf{Condition} &\textbf{Size} & \textbf{(per app), mins} & \textbf{(relative estimation)} \\ \hline
\endfirsthead
\multicolumn{9}{c}%
{\tablename\ \thetable\ -- \textit{Continued from previous page}} \\ \hline
\multirow{2}{*}{}%
\textbf{Article} &
\textbf{Artefact} &
\textbf{Testing} & \textbf{Testing} & \textbf{Search} & \textbf{Termination} &\textbf{Benchmark} & \textbf{Execution Time} & \textbf{Efficiency} \\ 
\textbf{Reference}&&\textbf{Approach} & \textbf{Technique} & \textbf{Algorithm} & \textbf{Condition} &\textbf{Size} & \textbf{(per app), mins} & \textbf{(relative estimation)} \\ \hline
\endhead
\hline 
\multicolumn{9}{r}{\textit{Continued on next page}} \\
\endfoot
\hline
\insertTableNotes
\endlastfoot

\rowcolor{mygray}\multirow{1}{*}{}
\cite{sigdroid} & 
SIG--Droid  & 
Code search--based &
Systematic & 
Depth--first  & Model exploration completeness &6& 3.5 (average) & \ccccc \\ \hline

\rowcolor{mygray}\multirow{2}{*}{}
\cite{patdroid} & 
PATDroid  & 
Code search--based &
Systematic  & 
Breadth--first  & Model exploration completeness &10& 4.3 (average) & \ccccc \\  \hline

\rowcolor{mygray}\multirow{2}{*}{}
\cite{ehbdroid} & 
EHBDroid  & 
Code search--based &
Systematic   & 
Modified Depth--first & Model exploration completeness &35& 10 (max) & \cccc  \\ \rowcolor{mygray}
&&&&(activity--directed)&&&& \\ \hline

\rowcolor{mygray}\multirow{2}{*}{}
\cite{xdroid} & 
Xdroid  & 
Code search--based  & 
Random & 
Uniform Random & Execution time &8& 30 (exact) & \cc \\ \rowcolor{mygray}
&&&&(based on Xmonkey)&&&& \\ \hline

\rowcolor{mygray}\multirow{2}{*}{}
\cite{sapienz} & 
Sapienz  & 
Multi-objective  & 
Hybrid\textcolor{blue}{*} & 
Multi-objective  & Execution time  &68& 60 (exact) & \ca \\ \rowcolor{mygray}
&&code search--based&&evolutionary search&&&&\\  \rowcolor{mygray}
&&&&(based on NSGA--II)&&&&\\ \hline

\rowcolor{mygray}\multirow{1}{*}{}
\cite{acteve} & 
ACTEve  & 
Code search--based & 
Systematic & 
Generational search & Depth of model exploration &5& 71 (average) & \ca \\  \hline

\rowcolor{mygray}\multirow{2}{*}{}
\cite{a3e} & 
A\textsuperscript{3}E  & 
Code search--based  & 
Targeted & 
Guide search & Model exploration completeness &28& 88 (average) & \ca \\ \rowcolor{mygray}
&(Targeted)&&&(based on taint--tracking)&&&& \\ \hline

\rowcolor{mygray}\multirow{2}{*}{}
\cite{collider} & 
Collider  & 
Code search--based &
Targeted & 
Breath--first & Model exploration completeness &5& 180 (min) & \ca \\  \hline

\rowcolor{mygray}\multirow{2}{*}{}
\cite{evodroid} & 
EvoDroid  & 
Code search--based  &
Systematic   & 
Evolutionary search & Model exploration completeness or &10& 3440 (average) & \ca \\  \rowcolor{mygray}
&&&&(step--wise segmented)&User--defined&&& \\ \hline

\multirow{2}{*}{}
\cite{orbit} & 
ORBIT  & 
GUI model--based &
Systematic  & 
Modified Depth--first & Model exploration completeness &8& 3.1 (average) & \ccccc \\ 
&&&&(forward--crawling)&&&&\\ \hline

\multirow{2}{*}{}
\cite{androframe} & 
AndroFrame  & 
GUI model--based  & 
Random  & 
Guided search & Execution time &100& 10 (exact) & \cccc \\  
&&&&(QLearning--based)&&&&\\ \hline

\multirow{1}{*}{}
\cite{droiddev} & 
DroidDEV & 
GUI model--based &
Systematic  & 
Best--first (informed search) & Model exploration completeness &20& 16 (average) & \ccc  \\ \hline

\multirow{2}{*}{}
\cite{mobolic} & 
Mobolic  & 
GUI model--based &
Systematic  & 
A* (informed search) & Model exploration completeness &10& 22 (average) & \cc \\ \hline

\multirow{2}{*}{}
\cite{mcrawlt} & 
MCrawlT  & 
GUI model--based & 
Systematic & 
Guided search & Model exploration completeness &30& 43 (average) & \ca \\  
&&&&(based on Backtracking)&&&&\\ \hline

\multirow{2}{*}{}
\cite{gat} & 
GAT  & 
GUI model--based & 
Systematic & 
Modified Depth--first & Model exploration completeness &9& 45 (average) & \ca \\  
&&&&(gesture-guided)&&&& \\ \hline

\multirow{1}{*}{}
\cite{droidwalker} & 
DroidWalker  & 
GUI model--based & 
Systematic & 
Depth--first & Execution time &20& 60 (exact) & \ca \\  \hline

\multirow{2}{*}{}
\cite{aimdroid} & 
AimDroid  & 
GUI model--based  & 
Systematic & 
Breadth--first  & Execution time &50& 60 (exact) & \ca \\  \hline

\multirow{2}{*}{}
\cite{a3e} & 
A\textsuperscript{3}E  & 
GUI model--based & 
Systematic & 
Depth--first & Model exploration completeness &28& 104 (average) & \ca \\
&(Depth--first)&&&&&&& \\  \hline

\multirow{1}{*}{}
\cite{autodroidcomb} & 
Autodroid   & 
GUI model--based  & 
Systematic & 
Combinatorial search &Execution time& 10 & 120 (exact) & \ca \\  
&(Combinatorial)&&&(based on greedy algorithm)&&&&\\ \hline

\multirow{1}{*}{}
\cite{autodroidcomb} & 
Autodroid   & 
GUI model--based  & 
Random & 
Modified Random &Execution time& 10 & 120 (exact) & \ca \\  
&(Frequency)&&&(frequency--based)&&&&\\ \hline

\multirow{2}{*}{}
\cite{stoat} & 
Stoat  & 
GUI  model--based &
Random  & 
Guided search & Execution time &93& 180 (exact) & \ca \\ 
&&&&(based on Markov Chain&&&& \\ 
&&&&Monte Carlo  sampling)&&&& \\ \hline

\multirow{2}{*}{}
\cite{swifthand} & 
SwiftHand  & 
GUI model--based & 
Systematic & 
Guided search & Execution time &10& 180 (exact) & \ca \\ 
&&&&(based on passive learning)&&&& \\ \hline

\multirow{1}{*}{}
\cite{guicc} & 
GUICC  & 
GUI model--based & 
Systematic & 
Breadth--first & Model exploration completeness &10& 180 (max) & \ca \\  \hline

\multirow{1}{*}{}
\cite{land} & 
LAND  & 
GUI model--based & 
Systematic & 
Breadth--first & Model exploration completeness &5& 180 (max) & \ca \\  \hline

\multirow{1}{*}{}
\cite{adautomation} & 
ADAutomation  & 
GUI model--based & 
Systematic & 
Depth--first & Depth of model exploration &2& 1170 (average) & \ca \\  \hline

\multirow{1}{*}{}
\cite{mobiguitar} & 
MobiGUITAR  & 
GUI model--based  & 
Systematic & 
Breath--first & Model exploration completeness &4& \textit{NA} & -- \\  \hline

\multirow{2}{*}{}
\cite{dynodroid} & 
Dynodroid  & 
GUI model--based  & 
Random & 
Biased Random & Number of events &50& \textit{N/A} & -- \\ 
&&&&(history--based)&&&& \\ \hline

\end{longtable}
\end{ThreePartTable}
\end{landscape}
}

{\fontsize{8}{10}\selectfont
\begin{landscape}
\begin{ThreePartTable}

\begin{longtable}{l|l|l|l|l|l|l}%
\caption{\textbf{Extracted data for RQ\#3 and its secondary questions: Practicality estimation.}
\label{tbl:14}} \\ \hline
\multirow{2}{*}{}%
\textbf{Article} &
\textbf{Artefact} &
\textbf{Code Coverage} & \textbf{Execution Time} & \textbf{Effectiveness} &\textbf{Efficiency}  & \textbf{Practicality} \\ 
\textbf{reference}&&\textbf{(average value), \%} & \textbf{(per app), mins} & \textbf{(relative estimation)}& \textbf{(relative estimation)} & \textbf{(relative estimation)} \\ 
\hline
\endfirsthead
\multicolumn{7}{c}%
{\tablename\ \thetable\ -- \textit{Continued from previous page}} \\ \hline
\multirow{2}{*}{}%
\textbf{Article} &
\textbf{Artefact} &
\textbf{Code Coverage} & \textbf{Execution Time} & \textbf{Effectiveness} &\textbf{Efficiency}  & \textbf{Practicality} \\ 
\textbf{reference}&&\textbf{(average value), \%} & \textbf{(per app), mins} & \textbf{(relative estimation)}& \textbf{(relative estimation)} & \textbf{(relative estimation)} \\ 
\hline
\endhead
\hline 
\multicolumn{7}{r}{\textit{Continued on next page}} \\
\endfoot
\hline
\endlastfoot

\cite{orbit} & 
ORBIT  & 
78 &
3.1 (average)   & 
\CCC & \ccccc & \cllll \\ \hline

\cite{sigdroid} & 
SIG--Droid  & 
78 &
3.5 (average)  & 
\CCC & \ccccc & \cllll \\ \hline

\rowcolor{mygray}\cite{droiddev} & 
DroidDEV  & 
91 &
16 (average)  & 
\CCCC & \ccc & \clll\hcl \\ \hline

\rowcolor{mygray}\cite{patdroid} & 
PATDroid  & 
60 &
4.3 (average)   & 
\CC & \ccccc & \clll\hcl \\ \hline

\cite{mobolic} & 
Mobolic  & 
92 &
22 (average)   & 
\CCCC & \cc & \clll \\ \hline

\cite{ehbdroid} & 
EHBDroid  & 
57 &
10 (max)  & 
\CC & \cccc & \clll \\ \hline

\rowcolor{mygray}\cite{evodroid} & 
EvoDroid  & 
81 &
3440 (average)  & 
\CCC & \ca & \cll \\ \hline

\cite{mcrawlt} & 
MCrawlT  & 
65 &
43 (average)   & 
\CC & \ca & \cl\hcl \\ \hline

\cite{stoat} & 
Stoat  & 
61 &
180 (exact)  & 
\CC & \ca & \cl\hcl \\ \hline

\cite{land} & 
LAND  & 
58 &
180 (max)  & 
\CC & \ca & \cl\hcl \\ \hline

\cite{droidwalker} & 
DroidWalker  & 
57 &
60 (exact)  & 
\CC & \ca & \cl\hcl \\ \hline

\multirow{2}{*}{}%
\cite{autodroidcomb} & 
Autodroid  & 
57 &
120 (exact)  & 
\CC & \ca & \cl\hcl \\ 
&(Combinatorial)&&&&&\\ \hline

\cite{sapienz} & 
Sapienz  & 
53 &
60 (exact)   & 
\CC & \ca & \cl\hcl \\ \hline

\multirow{2}{*}{}%
\cite{autodroidcomb} & 
Autodroid  & 
52 &
120 (exact)  & 
\CC & \ca & \cl\hcl \\ 
&(Frequency)&&&&& \\ \hline

\cite{xdroid} & 
Xdroid  & 
39 &
30 (exact)  & 
\C & \cc & \cl\hcl \\ \hline

\rowcolor{mygray}\cite{guicc} & 
GUICC  & 
43 &
180 (max)  & 
\C & \ca & \cl \\ \hline

\cite{dynodroid} & 
Dynodroid  & 
55 &
\textit{N/A}   & 
\CC & -- & -- \\ \hline

\cite{androframe} & 
AndroFrame  & 
\textit{NA} &
10 (exact)  & 
-- & \cccc & -- \\ \hline

\cite{gat} & 
GAT  & 
\textit{NA} &
45 (average)  & 
-- & \ca & -- \\ \hline

\cite{aimdroid} & 
AimDroid  & 
\textit{NA} &
60 (exact)  & 
-- & \ca & -- \\ \hline

\cite{acteve} & 
ACTEve  & 
\textit{NA} &
71 (average)  & 
-- & \ca & -- \\ \hline

\multirow{2}{*}{}%
\cite{a3e} & 
A\textsuperscript{3}E  & 
\textit{NA} &
88 (average)  & 
-- & \ca & -- \\ 
&(Targeted)&&&&&\\ \hline

\multirow{2}{*}{}%
\cite{a3e} & 
A\textsuperscript{3}E  & 
\textit{NA} &
104 (average)  & 
-- & \ca & -- \\* 
&(Depth--first)&&&&&\\ \hline

\cite{swifthand} & 
SwiftHand  & 
\textit{NA} &
180 (exact)  & 
-- & \ca & -- \\ \hline

\cite{collider} & 
Collider  & 
\textit{NA} &
180 (min)  & 
-- & \ca & -- \\ \hline

\cite{adautomation} & 
ADAutomation  & 
\textit{NA} &
1170 (average)  & 
-- & \ca & -- \\ \hline

\cite{mobiguitar} & 
MobiGUITAR  & 
\textit{NA} &
\textit{NA}  & 
-- & -- & -- \\ \hline

\end{longtable}
\end{ThreePartTable}
\end{landscape}
}

\section{Results}
\label{sec:results}
In this systematic review we specify 3 primary RQs regarding \textit{effectiveness}, \textit{efficiency}, and \textit{practicality} of the automated GUI testing techniques for mobile apps. In this section, we answer the primary RQs by discussing the extracted data from the primary studies. We use the resultant data which is collated and summarized  in Table~\ref{tbl:12}, Table~\ref{tbl:13}, and Table~\ref{tbl:14} in Section~\ref{subsec:datasynth}. 
 
\paragraph{Effectiveness} Effectiveness is one of the important characteristics of automated testing  techniques. For the automated testing techniques, effectiveness is usually assessed through the coverage which can be achieved upon completion of  tests execution. In particular, code coverage metric is useful means of effectiveness  assessment for the automated testing techniques. However, it is practically impossible to give an absolute estimation of effectiveness since all the testing techniques are evaluated on different data sets and testing environments. Thus, in this systematic review, we provide relative estimation of the effectiveness. To estimate the effectiveness, we unify the results of primary studies using code coverage metric, and averaging their code coverage results with respect to the data sets used for the experiments. 

We identify 4 conceptual characteristics of the automated testing tools which may affect effectiveness: \textit{model paradigm}, \textit{test generation approach}, \textit{test generation criteria}, and \textit{textual input generation}. From our observations of the extracted data, we conclude that test generation approach plays a dominant role in effectiveness. It innately identifies model paradigm and test generation criteria, which subsequently impact effectiveness of the testing technique. There is also 1 additional characteristic  such as textual input generation which affects effectiveness. Depending on the model paradigm, it may impact effectiveness to a different extent. For example, random-based and stochastic models are less likely to be affected due to their random nature, while deterministic models such as transition- or control-data-based could be affected severely.

From our given relevant estimation of effectiveness, we conclude that testing techniques which implement model-checking, symbolic execution, constraint solving, and search-based test generation approach tend to be more effective than those implementing random test generation. It could be explained by the fact that more sophisticated test generation approaches are likely to be more effective since they usually exploit the built  model heuristics to generate high coverage tests. 

We observe that random test generation approaches tend to use fault detection as a test generation criteria, while the others mainly focus on structural model or code coverage. It could be explained by the random nature of the testing approaches. In fact, randomly generated tests are likely to expose more bugs due to their unexpected (random) nature, while systematic ones may not hit the same number of bugs unless their coverage is 100\%. This is a main reason why systematic testing approaches focus on increasing structural model or code coverage aiming 100\%. Otherwise, random techniques will always take the first place being state-of-the-art and practice in the automated functional testing. 

We also observe that more sophisticated textual input generation mechanisms help to improve an effectiveness of the automated testing techniques. It could be explained by the fact that mobile apps have highly interactive nature, and thus they are crafted with mindset of being used by humans, not machines. As such, various app behaviours highly depend on the user inputs, textual in particular. So, currently, such app behaviours cannot be effectively tested by automated techniques due to the lack of adequacy in the automated textual input generation.

There are several techniques for which we were not able to identify their effectiveness due to an unavailability of the code coverage metrics which we are based on for the effectiveness estimation. However, we believe that their effectiveness could be the same or similar to those with equivalent characteristics. 

\paragraph{Efficiency} Efficiency is another important characteristic of automated testing techniques. For the automated testing techniques, efficiency is usually assessed through an execution time requiring the testing process to complete. However, efficiency of the automated testing techniques may vary depending on various factors such as experimental data sets, execution environment, computation power used for the experiment. As such, it is practically impossible to give an absolute estimation of the efficiency. So, in this systematic review, we give a relative estimation of the efficiency for the automated testing techniques. To estimate the efficiency, we extract from primary studies an execution time per app with respect to the data sets used for the  experiments. We estimate efficiency in minutes giving exact, average, minimum, or maximum time taken per app depending on the experimental set-up.

We identify 4 conceptual characteristics of the automated testing tools which may affect efficiency: \textit{testing approach}, \textit{testing technique}, \textit{search algorithm}, and \textit{termination condition}. From our observations of the extracted data, we conclude that testing approach plays a dominant role in efficiency. Another characteristic such as search algorithm  innately identifies testing technique which may subsequently  impact efficiency to a  different extent. For example, sophisticated search algorithms may require more complex implementation which eventually may slow down the overall efficiency of the testing technique, while simple techniques, such as random, are unlikely to impact the efficiency due to their simplicity. There is also 1 additional characteristic such as termination condition which affects the efficiency. Termination condition is innately identified by the testing technique. Depending on the testing technique, it can be automatically derived from the conditions of the constructed model during runtime. However, it can also be specified manually by a user to indicate when the testing procedure shall stop. For example, systematic testing techniques tend to use automatic termination condition, while random ones are usually rely on the user-specified  termination conditions.

From our given relevant estimation of efficiency,  we conclude that testing techniques which implement code search-based testing approaches tend to be more efficient than those implementing GUI model-based. It could be explained by the fact that code search-based testing approaches generate  tests being guided by more simple and compact models inferred from the app code rather than app GUI. In fact, complexity of the GUI models could be much higher, and their model size could be much larger than those  derived from the app code. 

We observe that systematic testing techniques tend to be less efficient than random ones. It could be explained by the fact that systematic techniques require more advanced search algorithms to implement a guided search, while random ones usually rely on uniform random. In contrast, systematic techniques use various heuristics of the inferred model to guide the testing process, while random ones do not require any heuristics so that they simply drive the testing process by sampling random actions from the  uniform distribution. 

We also observe that termination condition which relies on a user-specified condition is less efficient than the one which is automatically derived from the model properties. It could be explained by the fact that if the user specifies a termination condition, it usually implies that the testing approach is not capable of constructing a  high quality or deterministic model so that the termination condition cannot be derived from the model properties during runtime. However, if the termination condition is automatically derived from the model properties, it likely implies that the constructed models are of high quality and deterministic enough so that the model properties can be used to automatically determine when to stop the testing process. 

There are several techniques for which we were not able to identify their efficiency due to an unavailability of the execution time which we are based upon for the efficiency estimation, or simply because certain techniques do not specify termination condition via execution time, instead they specify number of events which are to be injected during the testing process. Once all the events have been injected, the testing process terminates. However, we believe that their efficiency could be the same or similar to those with equivalent characteristics.

\paragraph{Practicality}
Practicality is another important characteristic of automated testing techniques. 
In general, practicality depends on effectiveness and efficiency. As such, similar to effectiveness and efficiency, practicality may vary depending on the actual application of the technique. Innately, any user wishes to use an automated testing technique with high effectiveness and efficiency. However, it is practically impossible since there is no best single technique which is suited for all the testing purposes. 

Based on the estimated effectiveness and efficiency, we observe that the more effective a testing technique is, the less efficient it will be. It seems to be a trade-off between effectiveness and efficiency. So, the user should identify purpose of testing, and select most suitable testing technique for his particular purpose.
For example, if a target app has a highly complex GUI, it would be a better fit to use tools which implement (1) guided random, or/and (2) stochastic model-based GUI testing techniques which still can achieve an acceptable level of code coverage within a reasonable time. On the other hand, if the app has simple to medium GUI complexity, the user should choose one of the GUI model-based techniques which perform guided  systematic exploration, and commonly achieve high code coverage (of course, it depends on an actual inferred model [size, quality], and its handling) within a reasonable time. 

In practice, complex GUIs can also be tested by systematic GUI model-based testing techniques, not only random ones. However, for systematic exploration, their overall exercising time usually grows exponentially which is impractical in most of cases (unless the highest code coverage is a main test  target). Also, if app code and GUI are highly complex, the user may choose a tool which implements uniform random GUI testing technique since any other guided testing techniques may take very long time to complete, which could be impractical.

\section{Threats to Validity}
\label{sec:threats}
In this section we discuss main threats to validity of systematic reviews. We believe that such threats are to be minimized in order to increase a quality of the systematic review. So, we identify 5 threats which usually affect systematic reviews.

\paragraph{Search strategy}
Search strategy including search string and search resources may affect systematic review by not covering all possible primary studies conducted in the field. In practice, there is always a chance that certain primary studies may not be found. In fact, it is practically impossible to find all relevant to the topic of interest primary studies, however, we did our best to design such the search strategy which finds as many related articles as possible.

\paragraph{Publication bias}
To minimize possible publication bias, we conducted independent pilot review of the pre-selected primary studies to identify possible duplication of the results, proposed approaches, implemented techniques, ideas which may bias systematic review conclusion and results. The identified duplicates were excluded from this systematic review, however, we there could be a possibility that certain studies may still overlap giving a certain bias to our results and conclusions.

\paragraph{Human bias}
To minimize possible human bias, and have objective conclusions about primary studies reviews and their data quality, rather than subjective ones, we conducted systematic review with help of several people who are the authors of this systematic review. Having several individuals helps to avoid subjective personal opinions, thus driving our systematic review along the objective line  which is based on the facts and evidences described in the primary studies, but not personal opinions.

\paragraph{Quality assessment}
In our systematic review we intend to use only high quality primary studies, or at least with satisfactory quality. However, there is no defined metric which helps us to identify the level of quality of a primary study. So, to more subjectively assess the quality of the primary studies, we  developed our own criteria and scales. We believe that our estimation approach gives a sufficient ground to be trusted since there are also no evidences against. While we included in this systematic review only those studies which satisfy our selection criteria, and other were excluded, there could be a possibility that certain primary studies with satisfactory quality were missed.

\paragraph{Primary studies results validity}
During out study selection process, we found that many studies lack of discussion about threats to validity of their techniques and results. As such, we were not able to assess their validity. However to make it possible, we rely on 2 facts such as research design, and results discussion described in the primary studies. These 2 facts are commonly well-written and discussed, thus giving us a certain confidence about possible threats to the results validity. So, to deduce possible threats to validity based on the research design, and results discussion in the primary studies, we reply on our own knowledge in the topic. Such approach allows us to estimate how severe the threats could be so that we predict to which extent they may or may not impact the reported results.

\section{Conclusion}
\label{sec:conclusion}
From effectiveness, we observe that the existing automated testing techniques are not effective enough, and currently they achieve nearly half of the desired level of effectiveness. As such, there is still a gap which requires further research to improve existing techniques, or develop conceptually new test generation approaches to improve an effectiveness. In addition, we highlight another area for improvement such as automated textual input generation. Our observations show that most of the techniques currently use random text generation with may significantly impact the desired effectiveness, especially for mobile apps. However, automated relevant input generation is not a trivial task, this is why it is still an evolving field. 

From efficiency, we observe that current automated testing techniques are not efficient enough. In general, they provide medium-to-low efficiency requiring more than 30 minutes per app. Certain techniques even set several hours per app to more or less adequately explore app functionality. As such, there is still a large gap which requires further research to improve efficiency of the  automated testing techniques. In addition, we highlight another area for improvement focusing on the construction of compact yet high quality models, thus reducing their size and complexity. Existing approaches usually infer the models ``as are'', which commonly results in large and complex models. Even their quality is usually high, however  exploration time may grow exponentially to complete traversing such large models. The exploration time may take very long since existing search algorithms are not well-adapted to perform on such large models where exploration state-space could be infinite. 

From practicality, we observe that only nearly half of the existing tools could be used in practice, while the others are not practical due to their low effectiveness and efficiency. As such, most of the automated testing tools are not likely to be used in practice, while most probable practical tools may also lack high performance due to their performance gap which is induced by either low effectiveness or efficiency. So, there is still unresolved difficult practical problem which requires further investigations towards increasing practicality of the automated testing techniques by simultaneously improving their effectiveness and efficiency. 

\section*{References}

\end{document}